\documentclass[
  aps,
  prfluids,
  onecolumn,
  amsfonts,
  amssymb,
  amsmath,
  superscriptaddress,
]{revtex4-2}

\usepackage[utf8]{inputenc}
\usepackage{bm}  
\usepackage{ulem}  
\usepackage{siunitx}
\usepackage{booktabs}
\usepackage{graphicx}
\usepackage[svgnames]{xcolor}
\usepackage{adjustbox}
\usepackage{tabularx}
\usepackage{xspace}  
\usepackage{comment}
\usepackage{tikz} 

\usepackage[hidelinks]{hyperref}

\newcolumntype{Y}{>{\centering\arraybackslash}X}

\newcommand*{\vb}[1]{\boldsymbol{#1}}  
\newcommand*{\dd}{\mathrm{d}}  
\newcommand*{\pdv}[2]{\frac{\partial #1}{\partial #2}}
\newcommand*{\laplacian}{\nabla^2}
\newcommand*{\gradient}{\vb{\nabla}}

\newcommand*{\vvec}{\vb{v}}
\newcommand*{\kvec}{\vb{k}}
\newcommand*{\xvec}{\vb{x}}
\newcommand*{\fvec}{\vb{f}}
\newcommand*{\lvec}{\vb{l}}
\newcommand*{\rvec}{\vb{r}}
\newcommand*{\jvec}{\vb{j}}
\newcommand*{\fvecns}{\fvec_{\mathrm{ns}}}
\newcommand*{\phivecn}{\vb{\Phi}_\mathrm{n}}
\newcommand*{\phivecs}{\vb{\Phi}_\mathrm{s}}
\newcommand*{\nablavec}{\vb{\nabla}}
\newcommand*{\vvecn}{\vvec_\mathrm{n}}
\newcommand*{\vvecs}{\vvec_\mathrm{s}}
\newcommand*{\vortvec}{\vb{\omega}}

\newcommand*{\Cloop}{\mathcal{C}}
\newcommand*{\CircA}{\Gamma_{\!A}}

\newcommand*{\lambdaKolm}{\lambda^{\text{K41}}}
\newcommand*{\zetaKolm}{\zeta^{\text{K41}}}
\newcommand*{\Tlambda}{T_{\mathrm{\lambda}}}
\newcommand*{\sinc}{\mathrm{sinc}}
\newcommand*{\vtotvec}{\vb{v}_{\mathrm{tot}}}

\newcommand*{\krot}{k_{\mathrm{rot}}}
\newcommand*{\rhon}{\rho_\mathrm{n}}
\newcommand*{\rhos}{\rho_\mathrm{s}}
\newcommand*{\nun}{\nu_\mathrm{n}}
\newcommand*{\nus}{\nu_\mathrm{s}}
\newcommand*{\taylor}{\lambda_{\mathrm{T}}}

\newcommand*{\Rey}{\textit{Re}}
\newcommand*{\ReLambda}{\Rey_\lambda}
\newcommand*{\He}{$^4$He\xspace}
\newcommand*{\mean}[1]{\langle #1 \rangle}
\newcommand*{\Vnl}{V_{\mathrm{I}}}

\newcommand*{\NOTE}[1]{\textbf{\color{red}[#1]}}

\begin{document}

\title{Velocity circulation intermittency in finite-temperature turbulent superfluid helium}

\author{Nicol\'as P. M\"uller}
\affiliation{%
Université Côte d'Azur, Observatoire de la Côte d'Azur, CNRS,
Laboratoire Lagrange, Boulevard de l'Observatoire CS 34229 - F 06304 NICE Cedex 4, France
}
\author{Yuan Tang}
\author{Wei Guo}
\affiliation{%
  National High Magnetic Field Laboratory, 1800 East Paul Dirac Drive, Tallahassee, Florida 32310, USA
}
\affiliation{%
  Mechanical Engineering Department, FAMU-FSU College of Engineering, Florida State University, Tallahassee, Florida 32310, USA
}
\author{Giorgio Krstulovic}
\affiliation{%
Université Côte d'Azur, Observatoire de la Côte d'Azur, CNRS,
Laboratoire Lagrange, Boulevard de l'Observatoire CS 34229 - F 06304 NICE Cedex 4, France
}
\date{\today}

\begin{abstract}
We study intermittency of circulation moments in turbulent superfluid helium by using experimental grid turbulence and numerical simulations of the Hall-Vinen-Bekarevich-Khalatnikov model. More precisely, we compute the velocity circulation $\Gamma_r$ in loops of size $r$ laying in the inertial range. For both, experimental and numerical data, the circulation variance shows a clear Kolmogorov scaling $\mean{\Gamma_r^2} \sim r^{8/3}$ in the inertial range, independently of the temperature. Scaling exponents of high-order moments are comparable, within error bars, to previously reported anomalous circulation exponents in classical turbulence and low-temperature quantum turbulence numerical simulations.
\end{abstract}

\maketitle

\section{Introduction}

Turbulence, the disordered and chaotic motion of fluids, is an ubiquitous phenomenon in nature taking place at very different length scales, from astrophysical to micro scales \cite{Davidson2013}. Its dynamics is described by complex velocity fields dominated by vortices, regions of the flow with a strong local rotation. Despite great efforts and improvements made on its understanding over the last two centuries, 
there is still no full theory able to describe the dynamics of turbulent flows completely. 

The most traditional way of characterizing velocity fluctuations in classical turbulence (CT) at a given scale $r=|\rvec|$ is using the so-called structure functions $S_p(r) = \mean{\left[ \vvec(\xvec+\rvec) - \vvec(\xvec)\right]^p}$, where the brackets indicate an average in space, time or over different ensembles. When a large scale separation exists between the forcing scale $L$ and the dissipative scale $\eta$, the structure function displays power-laws as $S_p(r) \sim r^{\zeta_p}$ for $\eta\ll r \ll L$. For homogeneous isotropic flows, in 1941 Kolmogorov predicted the self-similar scaling $\zetaKolm_p = p/3$ (K41 prediction) \cite{Kolmogorov1941a}. Such a prediction is based on a mean-field approach and simply based on dimensional analysis.
Experiments and numerical simulations on homogeneous isotropic CT have however showed some deviations from K41 theory \cite{Frisch1995}. This breakdown of self-similarity is usually attributed to the highly intermittent nature of velocity fluctuations at small scales. 
There are several phenomenological theories based on multifractality intending to describe the intermittency of turbulent flows \cite{Kolmogorov1962,Benzi1984,She1994}. 

A different system with a manifest intermittency is quantum turbulence (QT), the turbulence taking place in superfluids \cite{Barenghi2014}. 
When liquid \He is cooled below the critical temperature of $\Tlambda=2.17$ K, it undergoes a phase transition into a superfluid state \cite{Pitaevskii2016}. Its dynamics at non-zero temperatures can be interpreted as a two-fluid system that mutually interact between themselves: a superfluid component with a velocity field $\vvecs$ that presents no viscosity, and a normal viscous component $\vvecn$ that is described by the classical Navier--Stokes equations \cite{Donnelly1991}. 
These two components can move in phase (coflow) or in counterphase (counterflow). 
In the first case, it has been observed both in experiments and numerical simulations that the statistical properties of the flow at large scales follows a behavior similar to classical fluids \cite{Maurer1998}. 
On the other hand, counterflow turbulence dynamics differs from classical fluids, displaying an inverse energy cascade and a breakdown of isotropy at small scales \cite{Biferale2019, Polanco2020a}.

In superfluid \He, the relative densities between the normal and superfluid components depend on temperature, and therefore there is an open discussion on whether there is or not a dependence of intermittency on the temperature. 
Experimental studies on QT at the wake of a disk in superfluid \He at temperatures between $1.3$ K $\leq T \leq \Tlambda$ show that there is no temperature dependence on the intermittency \cite{Rusaouen2017}. Other set of experiments on homogeneous isotropic QT show that there is no temperature dependence up to $p=6$, but there are some deviations from CT \cite{Varga2018, Tang2020}.
Numerical simulations on QT using different models like the Gross-Pitaevskii equation, shell-models or the HVBK (Hall-Vinen-Bekarevich-Khalatnikov) equations show a clear temperature dependence that is amplified at intermediate temperatures of $1.8 \leq T \leq 2$ K, where the density fractions of each component approach each other \cite{Krstulovic2016, Biferale2018a}. However, some HVBK-based shell models show an enhancement of intermittency on this temperature range while others show some decrease or even a non-intermittent behavior \cite{Boue2013,Shukla2016}.
Given the lack of consensus between experiments and numerical simulations, further studies are required on this subject. 

An alternative way of studying intermittency in turbulent flows is using moments of the velocity circulation instead of the velocity increments \cite{Migdal2020,Sreenivasan1995,Cao1996,Benzi1997}. The velocity circulation around a closed loop $\Cloop$ enclosing an area $A$ is defined by 

\begin{equation}
  \label{eq:circulation_intro}
  \CircA(\Cloop; \vvec) = \oint_\Cloop \vvec \cdot \dd\lvec = \iint_A \vortvec \cdot \vb{n} \, \dd S,  
\end{equation}

\noindent where in the second equality we make use of the Stokes theorem, with $\vortvec = \gradient \times \vvec$ the vorticity field. 
First theoretical studies on the statistics of velocity circulation suggested that the probability density function (PDF) follow the area rule, that is, within the inertial range of scales, they depend only on the minimal area circumscribed by the closed loop \cite{Migdal2020,Iyer2021}. Further numerical studies at low Reynolds numbers suggested that velocity circulation is a highly intermittent quantity, as well as velocity increments \cite{Sreenivasan1995,Cao1996,Benzi1997}. These results were also observed in experiments of homogeneous and isotropic turbulence in classical fluids \cite{Zhou2008}.
It was recently shown using high-resolution numerical simulations of the Navier--Stokes equations that the moments of circulation present a clear scaling 
\begin{equation}
  \mean{\Gamma_r^p} \sim r^{\lambda_p},\, \textrm{for} \quad \eta\ll r \ll L
\end{equation}
which deviate from Kolmogorov-based prediction $\lambdaKolm_p = 4p/3$ for larger moments \cite{Iyer2019}. Moreover, in numerical simulations of the Gross-Pitaevskii equation, a model for low-temperature superfluids, it was observed a very similar behavior between CT and QT \cite{Muller2021}.
The advantage of using the velocity circulation to study intermittency is that it is an integral quantity, and it allows for the development of new theories for intermittency \cite{Polanco2021a, Moriconi2021a}.
To our knowledge, there is still no experimental studies in superfluid \He of the scaling laws of velocity circulation.

In this work, we study intermittency of superfluid \He from the point of view of velocity circulation. 
We measure the circulation scaling in experiments of grid turbulence in superfluid \He and compare them with numerical simulations of the coarse-grained HVBK equations at different temperatures (see Sec.~\ref{sec:exp_and_model} for details on the experimental and numerical methods). The analysis is performed for large-scale statistics of QT. This article is organized as follows. In Sec.~\ref{sec:exp_and_model} we provide details of our experimental and numerical methods, including the model we use and the algorithm for computing the velocity circulation. Then, in Sec.~\ref{sec:results} we present our experimental and numerical results. Finally, in Sec.~\ref{sec:discussion} we summarize our results and discuss about known results on the circulation intermittency.

\section{Experimental setup and mathematical model}%
\label{sec:exp_and_model}

\subsection{Experimental setup}
\label{subsec:experimental_setup}

\begin{figure}[t]
	\centering
    \begin{tikzpicture}
      \node at (-9,8.8) {\textbf{(a)}};
      \node[above left] 
      {\includegraphics[width=0.51\textwidth]{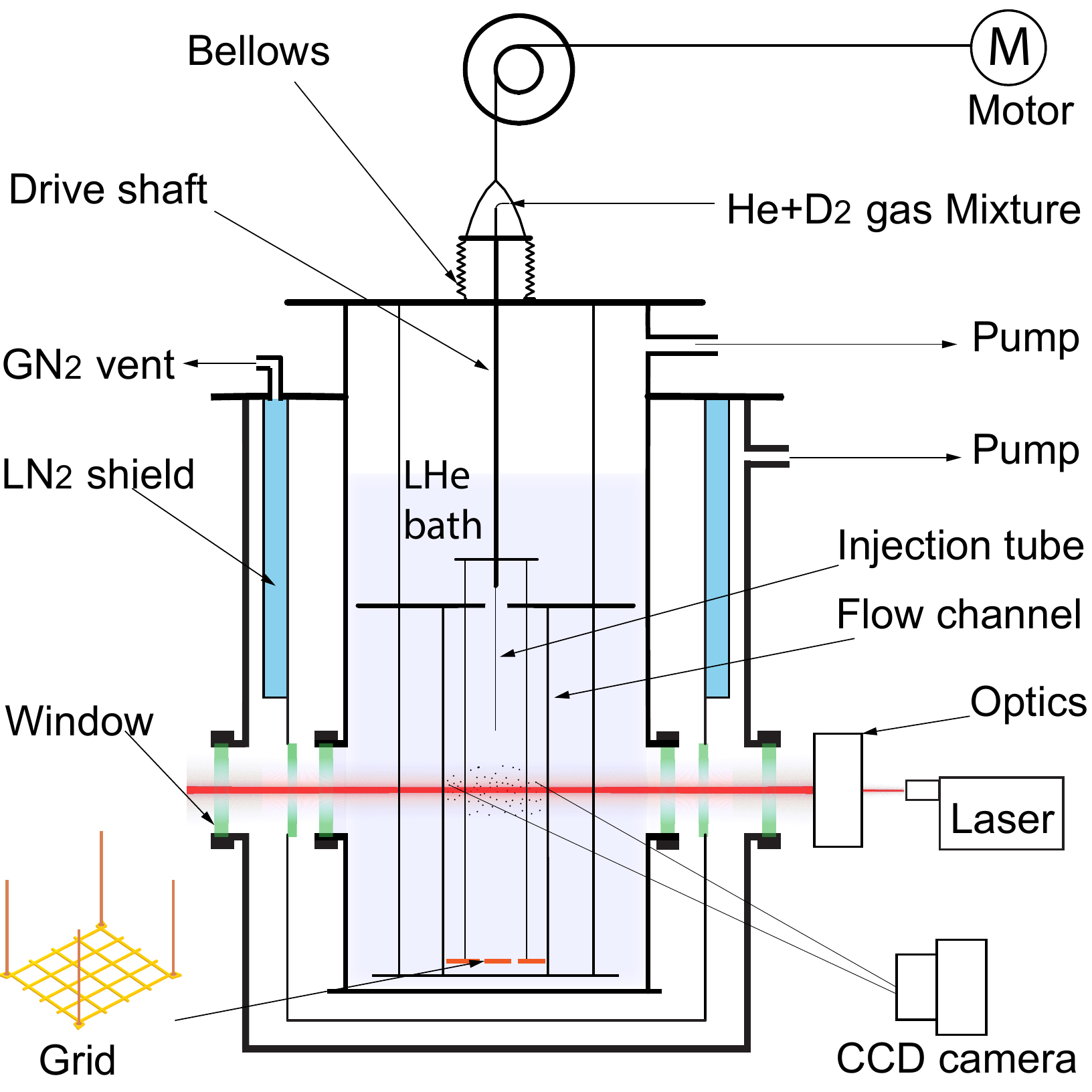}} ;
      \node[above right] 
      {\includegraphics[width=0.48\textwidth]{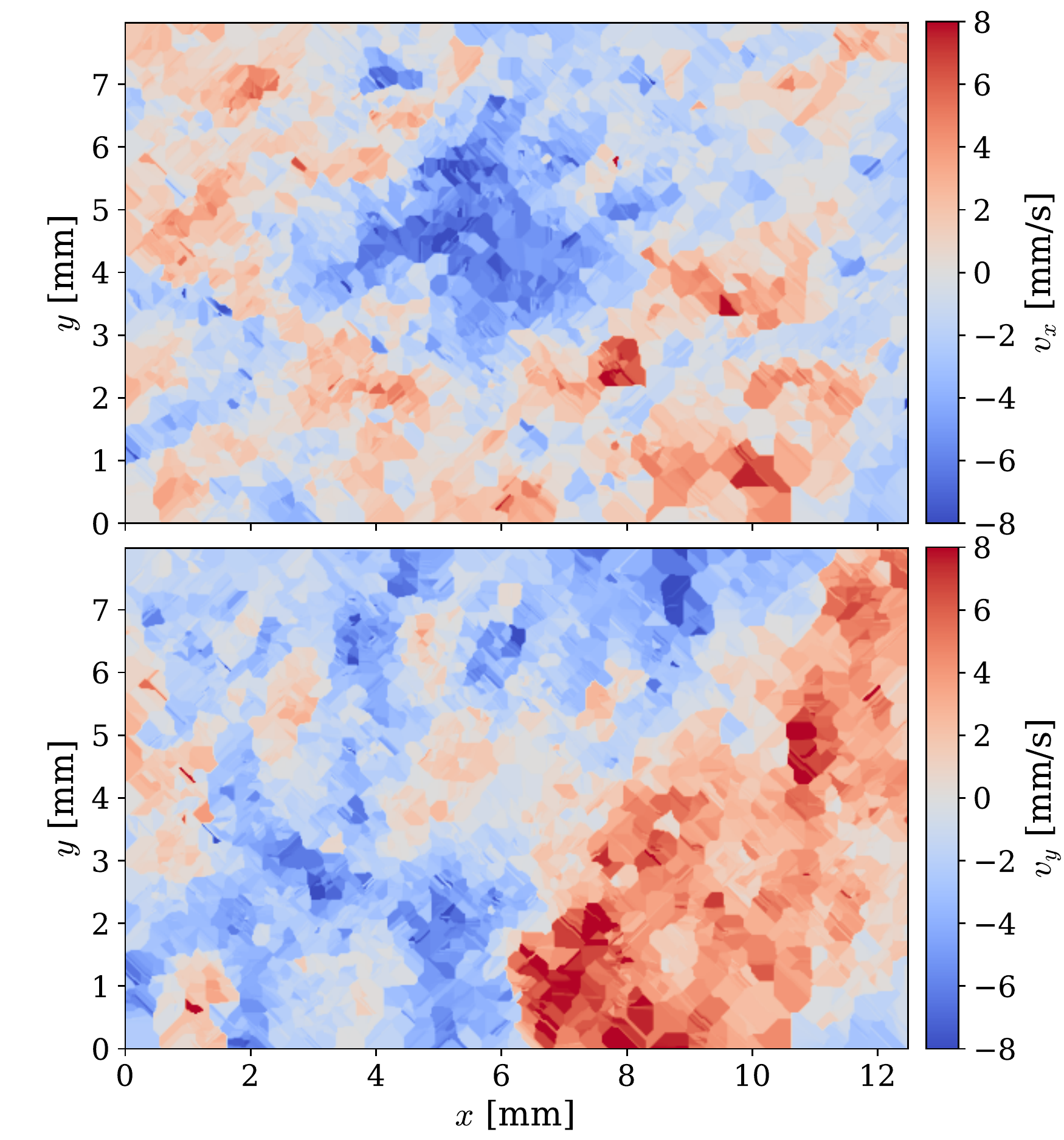}} ;
      \node at (.5,8.8) {\textbf{(b)}};
    \end{tikzpicture}
	\caption{(a) Experimental apparatus for grid turbulence of the superfluid helium-4. (b) Typical experimental velocity field obtained from the PTV measurements of grid turbulence in superfluid \He following the procedure described in Sec.~\ref{subsec:experimental_setup}.}
	\label{fig:exp_schematic}
\end{figure}

To examine the circulation statistics experimentally, we have conducted velocity-field measurements in quasiclassical turbulence generated in He II by a towed grid using the particle tracking velocimetry (PTV) method~\cite{Tang2020}. The experimental apparatus, shown in Fig.~\ref{fig:exp_schematic} (a), consists of a transparent cast acrylic flow channel with a cross-section area of $1.6\times1.6$ cm$^2$ and a length of $33$ cm immersed vertically in a He II bath (more details of the setup can be found in Ref.~\cite{Mastracci2018}). 
The bath temperature is controlled by regulating the vapor pressure. A brass mesh grid with a spacing of $3$ mm and $40\%$ solidity is suspended in the flow channel by stainless-steel thin wires at the four corners. A linear motor outside the cryostat can pull the wires and hence the grid at a speed up to $60$ cm/s. In the current work, we used a fixed grid speed at $30$ cm/s. To probe the flow, we adopt the PTV method using solidified D$_2$ tracer particles with a mean diameter of about $5$~$\mu$m~\cite{Mastracci2018}. These particles are entrained by the viscous normal-fluid flow due to their small sizes and hence small Stokes number~\cite{Tang2020,Mastracci2018a,Polanco2020}, but they can also get trapped on quantized vortices in the superfluid~\cite{Mastracci2019,Tang2021,Giuriato2019,Giuriato2020}. A continuous-wave laser sheet (thickness: $200$~$\mu$m, height: $9$ mm) passes through the center of two opposite side walls of the channel to illuminate the particles. The motion of the particles is then capture by a CCD camera at 200 frames per second at an angle perpendicular to both the flow channel and the laser sheet. We set $t=0$ when the grid passes the center of the view window and typically record the particle motion continuously for $40$ s. A modified feature-point tracking routine~\cite{Mastracci2018} is adopted to extract the trajectories of the tracer particles from the sequence of images. In the current work, we focus on analyzing the data obtained in the time interval $t=3$ s to $t=5$ s at two bath temperatures, i.e., $T=1.65$ K and $1.95$ K. The turbulence at these decay times appears to be reasonably homogeneous and isotropic, and its turbulence intensity is relatively high such that an inertial range exists~\cite{Tang2020}.

For circulation analysis, it is more convenient to have two-dimensional Eulerian velocity field. In order to generate this information using the spatially sparse PTV data, we adopt the method reported in Ref.~\cite{Tang2021a}. We first combine the velocity data $\vvec(x,y)$ obtained from 11 successive images into a single velocity-field image. This procedure assumes that during the acquisition time of these 11 images (i.e., 50 ms), the velocity field does not change considerably so that these data describe a single instantaneous velocity field. Then, we divide the combined image into square cells with side length $\Delta=0.02$ mm so most of the cells have at least 1-2 data points. The velocity assigned to the center of each cell is calculated as the Gaussian-averaged velocity of particles inside the cell with a Gaussian-profile variance $\sigma\approx \Delta/2$ to ensure that the Gaussian weight drops to near zero at the cell's edge. Occasionally, there may not be any particles that fall inside a particular cell. In this case, we increase the size of this cell by a factor of two, and this process may be repeated until a few particles fall in the enlarged cell so that the velocity at the cell center can be determined. A representative resulted velocity field $\vvec(x,y)$ obtained at $T=1.95$~K is shown in Fig.~\ref{fig:exp_schematic} (b).

\subsection{Model for superfluid helium}
\label{subsec:models}

The dynamics of superfluid helium at finite temperatures and scales larger than the inter-vortex
distance can be described by the coarse-grained Hall-Vinen-Bekarevich-Khalatnikov (HVBK) equations \cite{Donnelly1991,Boue2013,Biferale2017,Polanco2020}

\begin{gather}
  \frac{\partial \vvecn}{\partial t} + \vvecn \cdot \nablavec \vvecn = 
    - \frac{1}{\rho_n} \nablavec p_n + \nu_\mathrm{n} \laplacian \vvecn - \frac{\rho_s}{\rho_n} 
    \fvecns + \phivecn, \label{eq:HVBK_n} \\
  \frac{\partial \vvecs}{\partial t} + \vvecs \cdot \nablavec \vvecs = 
    - \frac{1}{\rho_s} \nablavec p_s + \nu_\mathrm{s} \laplacian \vvecs +  \fvecns + 
    \phivecs, \label{eq:HVBK_s} \\
  \nablavec \cdot \vvecn = \nablavec \cdot \vvecs = 0. \label{eq:HVBK_inc} 
\end{gather}

\noindent This incompressible two-fluid model describes the motion of the normal ($\vvecn$) and
superfluid ($\vvecs$) components via two coupled Navier--Stokes equations. The kinematic viscosity
is related to the dynamic one via $\nun = \mu / \rhon$, $p_{\mathrm{n,s}}$ is the hydrodynamic
pressure of each component, and the total density of the fluid is $\rho = \rhon + \rhos$. The
superfluid component also dissipates via an effective viscosity $\nus$ that takes
into account dissipative effects taking place at small scales that the HVBK model is not able to
resolve, like quantum vortex reconnections and Kelvin waves \cite{Koplik1993,Bewley2008,Villois2020, Krstulovic2012,Fonda2014}. Both Navier--Stokes equations are coupled through the mutual friction force
between both velocity components $\fvecns = \alpha \Omega_0 (\vvecn - \vvecs)$, with
$\alpha=\alpha(T)$ the mutual friction coefficient that depends on the temperature of the system.
The frequency $\Omega_0$ is proportional to the vortex line density and to the quantum of
circulation of the superfluid, and is estimated as $\Omega_0^2 = \mean{|\vortvec_s|^2} / 2$ with
$\vortvec_s = \nablavec \times \vvecs$ the superfluid vorticity and $\left< . \right>$ denoting a
spatial average \cite{Biferale2018a, Polanco2020}.  We use two independent large-scale Gaussian
random forces $\phivecn(\xvec)$ and $\phivecs(\xvec)$ to excite both fluid components and obtain a
stationary state.

We study the scaling of velocity circulation in superfluid \He at different temperatures by solving
numerically the HVBK equations \eqref{eq:HVBK_n}-\eqref{eq:HVBK_inc} using a fully dealiased Fourier
pseudospectral code in a periodic cubic domain and a third-order Runge-Kutta integration in time
(See Ref. \cite{Homann2009} for details). We perform 7 numerical simulations of these equations for
temperatures that vary between $T=1.3$ K and $T=2.1$ K, using $N=1024$ linear collocation points in
each direction. 
All the parameters used for each numerical simulation are shown in Table~\ref{tab:runs}.  We
extracted the values of the effective superfluid viscosity from \citet{Boue2015a}.  

\begin{table}[t]
\begin{tabularx}{\textwidth}{ X Y Y Y Y Y Y Y Y}
  \hline \hline
  RUN & N & $T$ (K) & $\alpha$  & $\rho_s/\rho$ & $\rho_n/\rho$ & $\nu_s / \nu_n$ & $\ReLambda^n$ & $\ReLambda^s$ \\
  \hline
  I   & 1024 & 1.3  &  0.034   &  0.952   &  0.048   & 0.043 & 34  & 412  \\
  II  & 1024 & 1.5  &  0.072   &  0.889   &  0.111   & 0.2   & 187 & 651  \\
  III & 1024 & 1.79 &  0.156   &  0.696   &  0.304   & 0.8   & 358 & 427  \\
  IV  & 1024 & 1.9  &  0.206   &  0.574   &  0.426   & 1.25  & 500 & 419  \\
  V   & 1024 & 1.96 &  0.244   &  0.504   &  0.496   & 1.50  & 550 & 410  \\
  VI  & 1024 & 2.05 &  0.347   &  0.362   &  0.638   & 1.87  & 550 & 345  \\
  VII & 1024 & 2.1  &  0.481   &  0.259   &  0.741   & 2.5   & 406 & 193  \\
  \hline \hline
\end{tabularx}
\caption{ Table of parameters for the numerical simulations of the HVBK equations. $N$ corresponds to the linear resolution on each direction,
$T$ is the temperature of the HVBK system expressed in Kelvin units, $\alpha$ the mutual friction coefficient, $\rho_s$ and $\rho_n$ the 
superfluid and normal densities, respectively, $\nu_s/\nu_n$ the ratio of the kinematic viscosities, and $\ReLambda^{n,s}$ to the Taylor-microscale Reynolds number $\ReLambda=v_{\rm rms} \lambda_{\rm T} / \nu$.}
\label{tab:runs}
\end{table}

For comparison, we also use data from reference \cite{Muller2021,Polanco2021a}. In particular, we use the circulation exponents of classical turbulence obtained by integrating the Navier-Stokes equations with a Taylor-microscale Reynolds number of $\ReLambda=510$, and zero-temperature quantum turbulence generated by using the Gross--Pitaevskii (GP) model, with a separation between the integral length scale $L_{\rm I}$ and the healing length $\xi$ of $L_{\rm I}/\xi=820$. Using the inter-vortex distance $\ell$ as the equivalent of the Taylor microscale in the GP model, we can obtain a microscale Reynolds number of $\ReLambda^{\rm GP} \equiv 15 L_{\rm I} / \ell = 440$. In both cases, the numerical simulations have a linear spatial resolution of $N=2048$.

\subsection{Data analysis}
\label{subsec:analysis}

The velocity circulation for the HVBK numerical simulations is computed using the Fourier
coefficients of the velocity fields of each component using our openly available code
\cite{Circulation}. Over each two-dimensional $L$-periodic slab of the system, in the three
different orientations, we compute the circulation over square loops of different sizes $r$ centered
at each point $\xvec = (x,y)$ of the domain as the convolution \cite{Polanco2021a}

\begin{equation}
  \Gamma_r(\xvec) = \int_{S_r(\xvec)} \omega_{\mathrm{n,s}}(\xvec') \dd^2 \xvec' = \int \int_{} H_r(\xvec - \xvec')
  \omega_{\mathrm{n,s}}(\xvec') \dd^2 \xvec,
  \label{eq:circulation_method}
\end{equation}

\noindent where $\omega_{\mathrm{n,s}} = (\nablavec_{\mathrm{2D}} \times \vvec_{\mathrm{n,s}})\cdot
\hat{z}$ is the two-dimensional vorticity of the normal or superfluid component for each slab and
$S_r(\xvec)$ a squared planar surface of linear size $r$ centered at $\xvec$. The convolution kernel
is defined as $H_r(\xvec)=\Pi(x/r)\Pi(y/r)$, where $\Pi(x)=1$ for $|x|<1/2$ and 0 otherwise, so that
it can be written in Fourier space in terms of the normalized sinc function as $\hat{H}_r(k_x, k_y)
= (r/L)^2 \sinc(k_xr/2\pi)\sinc(k_yr/2\pi)$. 

This method can be used to compute the simulations for the normal and superfluid components obtained from the numerical simulations of the HVBK equations due to their periodicity. However, the velocity fields obtained from experiments are not
periodic. Therefore, instead of using the Fourier coefficients, we compute the circulation directly
from the velocity field following the first equality in Eq.~\eqref{eq:circulation_intro}.

\section{Results}
\label{sec:results}

\subsection{Low order statistics from experimental data}
\label{subsec:experimental_results}

\begin{figure}[t]
  \centering
  \includegraphics[width=1\textwidth]{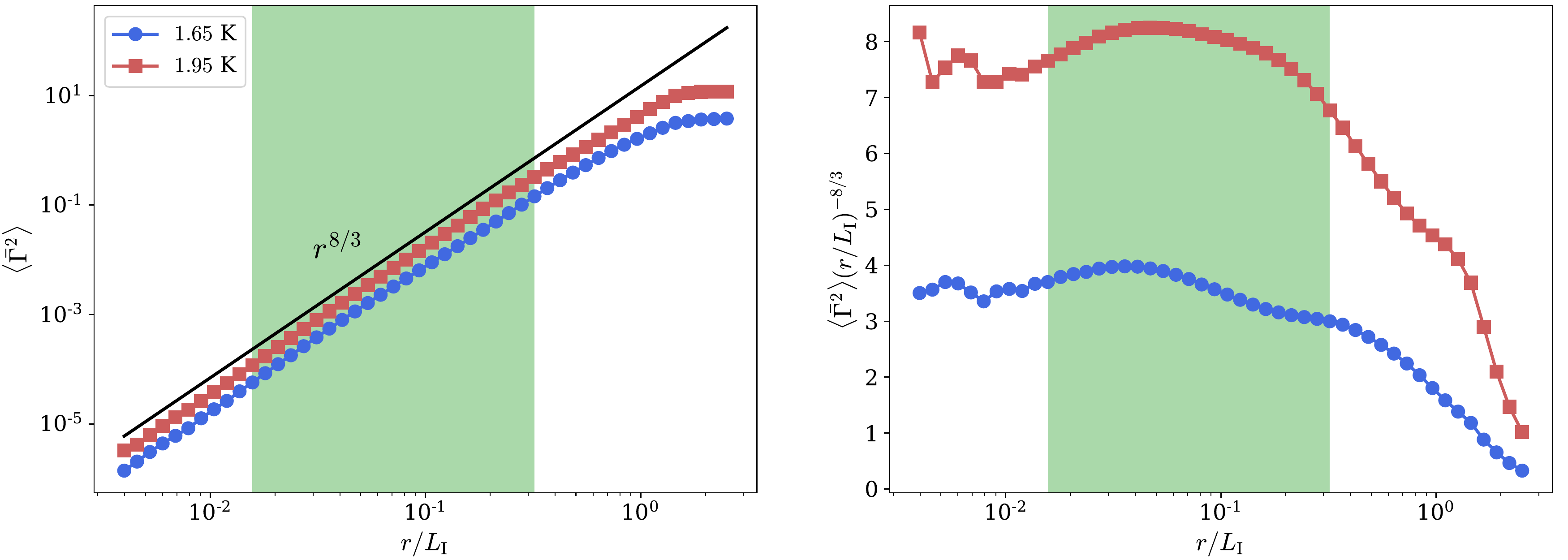}
  \caption[]{%
    Left panel: Circulation variance for the average of the experimental acquisitions at temperatures $T=1.65$ K (blue circles) and $T=1.95$ K (red squares). The green shaded area indicates the inertial range. The black solid line corresponds to Kolmogorov scaling law $r^{8/3}$. Right panel: Circulation variance compensated by Kolmogorov scaling. 
    }
    \label{fig:exp_variance}
\end{figure}

We analyse the data obtained from several realizations of grid turbulence in superfluid helium at temperatures $T=1.65$ K and $T=1.95$ K following the experimental setup described in Sec. \ref{subsec:experimental_setup}. 
We determine that the system reaches a regime of fully developed turbulence between three and five seconds after the grid passes through the center of the region. We study and obtain a two-dimensional Eulerian velocity field every $0.1$ s within this time interval on a rectangular window of $7.98 \times 12.48$ mm, following the procedure described in Sec.~\ref{subsec:experimental_setup}. These velocity fields allow us to compute the velocity circulation around squared planar loops of different linear sizes $r$, as described in Sec.~\ref{subsec:analysis}. 
As the velocity field is not periodic, we analyse a reduced window of $(L_x-r, L_y-r)$, obtaining a reduced amount of statistics for larger loops.
The mean energy injected on the system may fluctuate between the different realizations of the flow, so we normalize the circulation as $\bar{\Gamma} = \Gamma / \Gamma_0$ with $\Gamma_0 = v_\mathrm{rms} L_{\rm I}$ the circulation at large scales. In this manner, we are able to compare and average all the different realizations, where $L_{\rm I} = \int k^{-1}E(k) dk / \int E(k) dk$ is the integral length scale of the flow, $E(k)$ the energy spectrum, and $v_\mathrm{rms} = \sqrt{(v_x^2 + v_y^2)/2}$ the root-mean-square velocity of each flow realization. The typical integral length scale in our experiments is $L_{\rm I} = 4.5$ mm and the typical root-mean-square velocity is $v_{\rm rms} = 1.7$ mm/s.
Figure \ref{fig:exp_variance} shows the circulation variance of the averaged measurements. In the inertial range, represented by the green-shaded region, the circulation variance follows a scaling that approximates Kolmogorov one $\lambdaKolm_2 = 8/3$ for both temperatures. Moreover, when the variances are compensated by $\lambdaKolm_2$, they approach to a plateau.


\begin{figure}[t]
  \centering
  \includegraphics[width=1\textwidth]{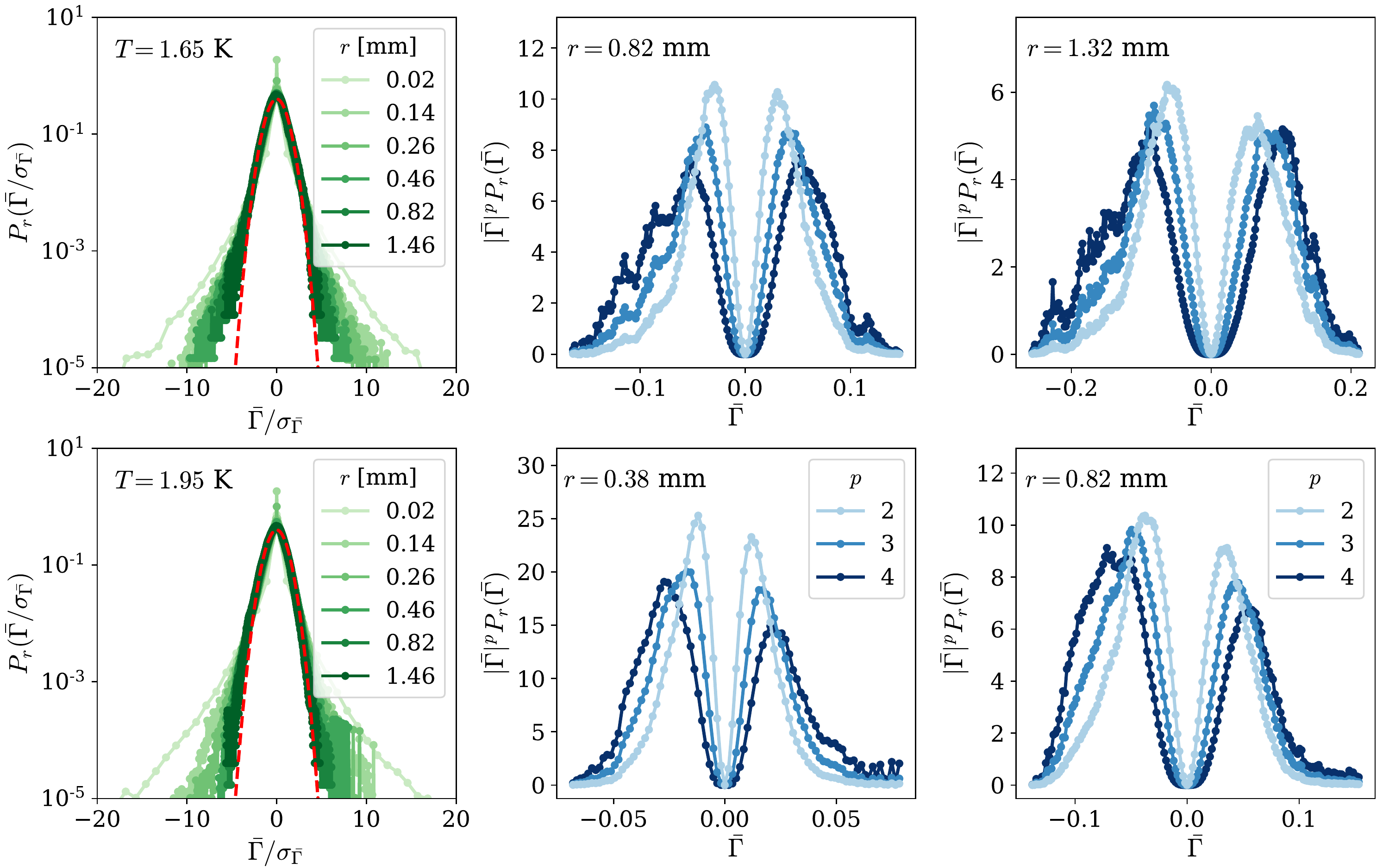}
  \caption[]{%
    Left panel: Experimental PDFs of the velocity circulation at different length scales. Each row corresponds to a different temperature. Middle and right panels: Circulation integrands up to moment four for two different length scales laying within the inertial range. The statistical convergence starts to fail on the fourth moment.}
    \label{fig:exp_integrands}
\end{figure}

Figure \ref{fig:exp_integrands} shows the probability density functions (PDFs) of the velocity circulation for both temperatures and for different loop sizes (in green). At small scales, the PDFs present heavy tails, a clear signature of intermittency. As the size of the loop increases, they collapse and approach to a Gaussian distribution (red dashed line). This behavior is similar to the one observed for the velocity circulation in numerical simulations of the Navier--Stokes and the Gross--Pitaevskii equations \cite{Iyer2019,Muller2021,Polanco2021a}, and experiments in classical fluids \cite{Zhou2008}. 

The study of high-order moments of the circulation $\mean{\Gamma^p}$ usually requires a large amount of data for statistical convergence \cite{Anselmet1984}. 
Measured moments of order $p$ cannot be trusted if the integrands $\Gamma^p P_r(\Gamma)$, for a given length scale within the inertial range, do not go to zero for the largest measured value of $\Gamma$, since the assumption $\mean{\Gamma^p} = \int_{-\infty}^{\infty} \Gamma^p P_r(\Gamma) \dd \Gamma \approx \int_{-\Gamma_c}^{\Gamma_c} \Gamma^p P_r(\Gamma) \dd \Gamma$, with $\Gamma_c$ the circulation cut-off, breaks down. In Fig. \ref{fig:exp_integrands} we also show the circulation integrands of the experimental measurements up to the fourth-order (in blue) for length scales within the inertial range. In particular, for the highest order shown here, the tails fail to converge for some scales. This behavior suggests that, at best, moment of order four are borderline in terms of statistical converge. 

\begin{figure}[t]
  \centering
  \includegraphics[width=1\textwidth]{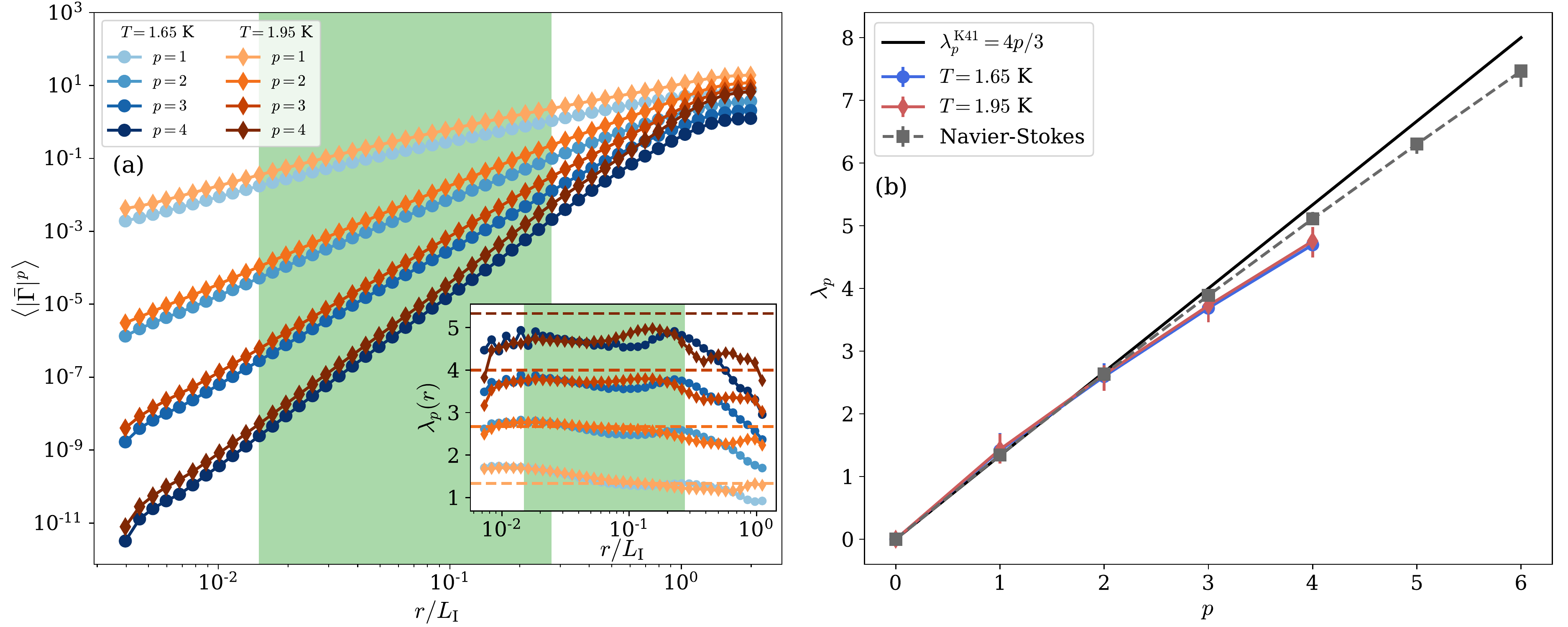}
  \caption[]{%
  (a) Circulation moments for the experimental data up to order four for temperatures $T=1.65$ K (blue circles) and $T=1.95$ K (orange diamonds). The inset shows the local slope of the circulation moments and Kolmogorov scaling as dashed horizontal lines. Green shaded areas indicate the inertial range for both temperatures. (b) Scaling exponents of the experimental measurements. The error bars indicate the maximum and minimum values of the local slope within the inertial range. As reference, we show Kolmogorov scaling $\lambdaKolm_p = 4p/3$ and the scaling exponents of classical turbulence obtained from numerical simulations of the Navier--Stokes equations (gray squares). }
    \label{fig:exp_scaling}
\end{figure}

The circulation moments up to the fourth order for $T=1.65$ K (blue circles) and for $T=1.95$ K (orange diamonds) are shown in Fig. \ref{fig:exp_scaling} (a). We compute the odd-order moments using the absolute value of the circulation.
The local slopes, defined as the logarithmic derivative $\lambda_p(r) = \dd \log \mean{|\Gamma|^p} / \dd \log r$, approach to a plateau within the inertial range, obtaining the scaling exponents $\lambda_p$ shown in Fig. \ref{fig:exp_scaling} (b). The error bars correspond to the maximum and minimum values of the local slopes in the inertial range. 
Up to the third order, the scaling exponents seem to follow Kolmogorov scaling law for the circulation $\lambdaKolm = 4p/3$. For higher orders, they start deviating from this prediction taking smaller values, and hence a stronger intermittency. As a reference, we show the scaling exponents of CT obtained from numerical simulations of the incompressible Navier--Stokes equations, taken from \cite{Polanco2021a}. 
Our experimental data starts deviating from the classical limit for increasing order moments.
However, data does not allow to enforce this claim due to a possible lack of statistics to compute the forth order moment, as shown in Fig.~\ref{fig:exp_integrands}. See Sec.~\ref{sec:discussion} for a further discussion.

\subsection{HVBK results}
\label{subsec:HVBK_results}

\begin{figure}[t]
  \centering
  \includegraphics[width=.49\textwidth]{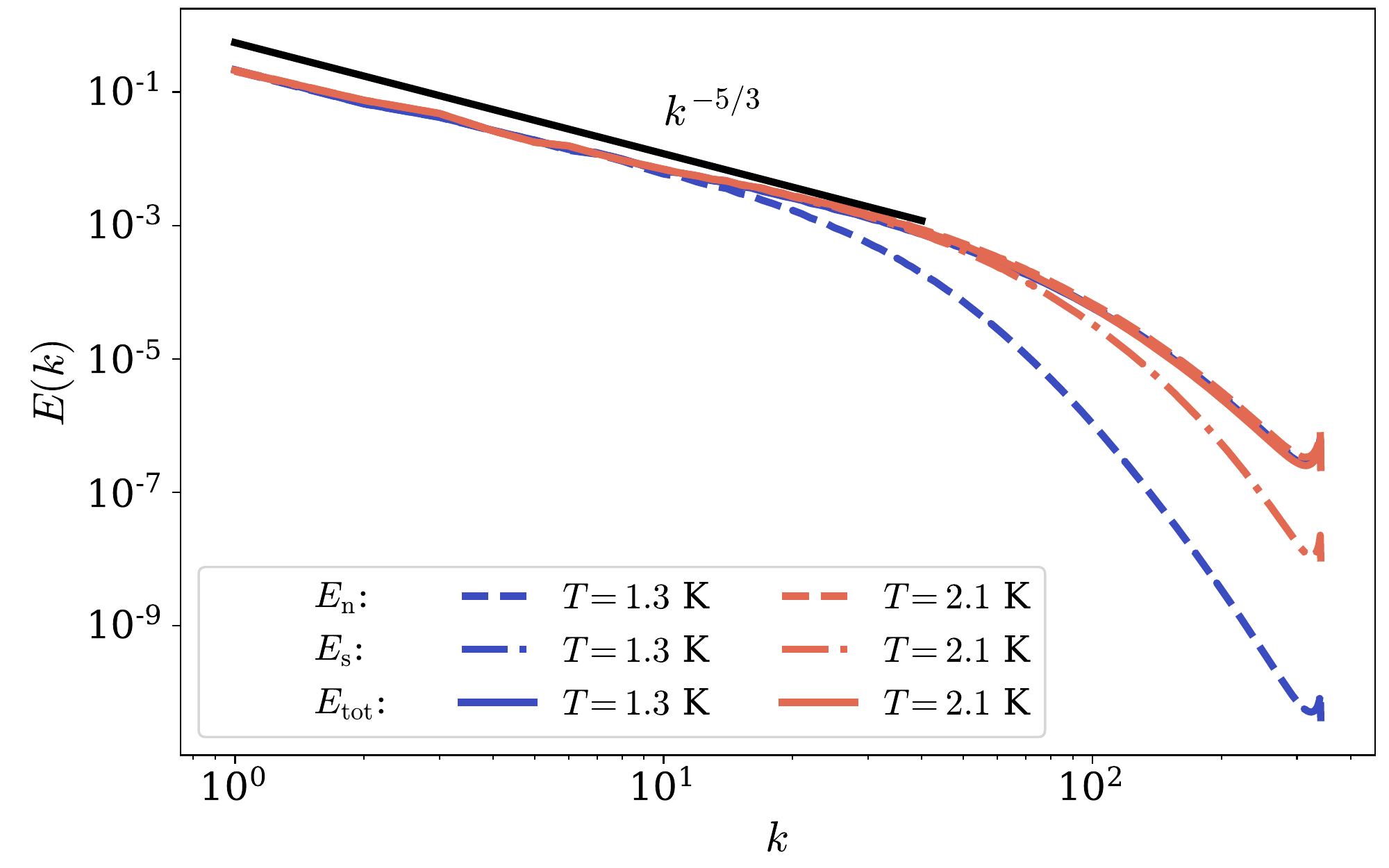}
  \includegraphics[width=.49\textwidth]{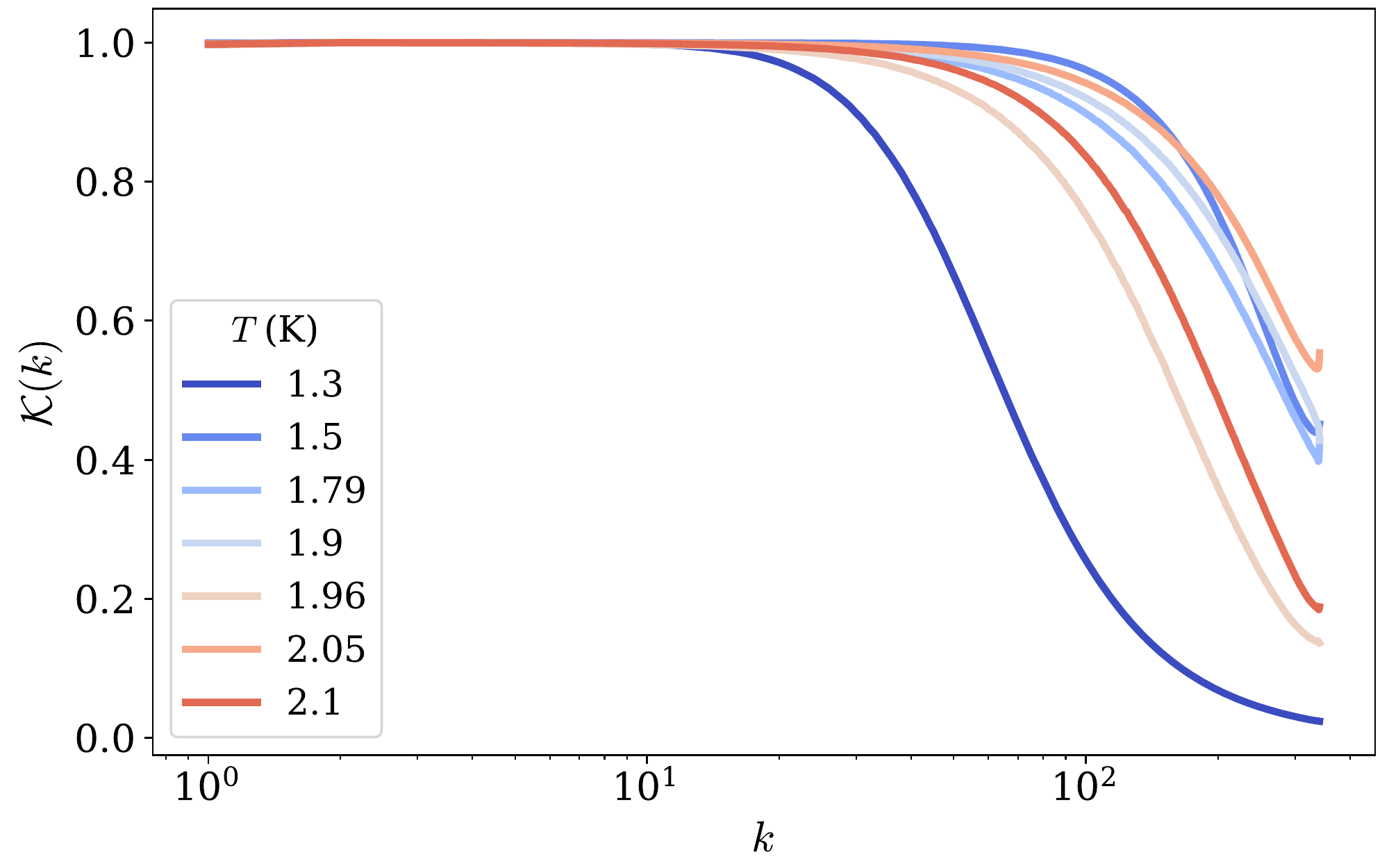}
  \caption[]{%
    Left panel: Spectra of the different energy components for the lowest and highest temperature of the simulations. Right panel: Velocity cross-correlation defined in Eq.~\eqref{eq:cross} for different temperatures. 
    }
    \label{fig:spectra}
\end{figure}

The experimental results presented in Sec.~\ref{subsec:experimental_results} provide a first evidence of circulation scaling in superfluid helium turbulence for low-order moments, in particular observing a Kolmogorov scaling up to the third order. However, the analysis of high-order moments cannot be completely trusted due to the lack of statistics. 
To provide an insight on this aspect, we perform numerical simulations of the coarse-grained HVBK equations \eqref{eq:HVBK_n}-\eqref{eq:HVBK_inc} using typical parameters for superfluid \He (see Table~\ref{tab:runs}). We force the system with two independent random forces to obtain a stationary state of homogeneous isotropic QT.
The two-fluid HVBK model describes the motion of the normal and the superfluid components at finite temperatures. Therefore, the turbulent properties of the flow may differ between them so each velocity component, in principle, should be studied independently. Figure \ref{fig:spectra} shows the energy spectra of each velocity component in a statistically steady turbulent regime for two temperatures, the highest and the lowest ones studied in this work. For both temperatures and velocity components, the energy spectra display a scaling close to Kolmogorov one $E_{\mathrm{n,s}} \sim k^{-5/3}$ within an inertial range that varies depending on the temperature and the velocity component. The reason for these variations is that the normal and effective superfluid viscosities vary, and also present a different temperature dependence. One way of defining a homogeneous inertial range to facilitate the analysis of these two velocity components is by studying the total velocity field $\vtotvec = \jvec / \rho$ with $\jvec = \rhos \vvecs + \rhon \vvecn$ the total momentum density. 

The use of the total velocity could be valid under the assumption of \textit{locking} between both velocity components, in the sense of $\vvecn \approx \vvecs$ \cite{Vinen2002, Boue2015a}. 
One way of quantifying the scale-by-scale locking is with the velocity cross-correlation \cite{Boue2013, Biferale2018a} 

\begin{equation}
  \mathcal{K}(k) = \frac{2E_\mathrm{ns}(k)}{E_\mathrm{n}(k) + E_\mathrm{s}(k)},
  \label{eq:cross}
\end{equation}

\noindent with $E_\mathrm{ns}(k)$ the cross-velocity energy spectrum associated to $\vvecn \cdot \vvecs$. If the cross-correlation is equal to one, it indicates that both components are completely locked while if it approaches to zero the superfluid and normal velocities are statistically independent. Figure \ref{fig:spectra} (b) shows that for all temperatures the velocity components are locked with $\mathcal{K}(k) > 0.95$ at least up to $k \approx 50$ except for the lowest temperature case $T=1.3$ K, where the locking stops at $k \approx 20$ as a consequence of the small proportion of normal density. In the inertial range, where the energy spectrum obeys Kolmogorov scaling, both fluid components are locked, so the study of the normal, superfluid or total velocities should be statistically equivalent. Therefore, most of the following analysis on the velocity circulation is done using the total velocity. 

\begin{figure}[t]
  \centering
  \includegraphics[width=1\textwidth]{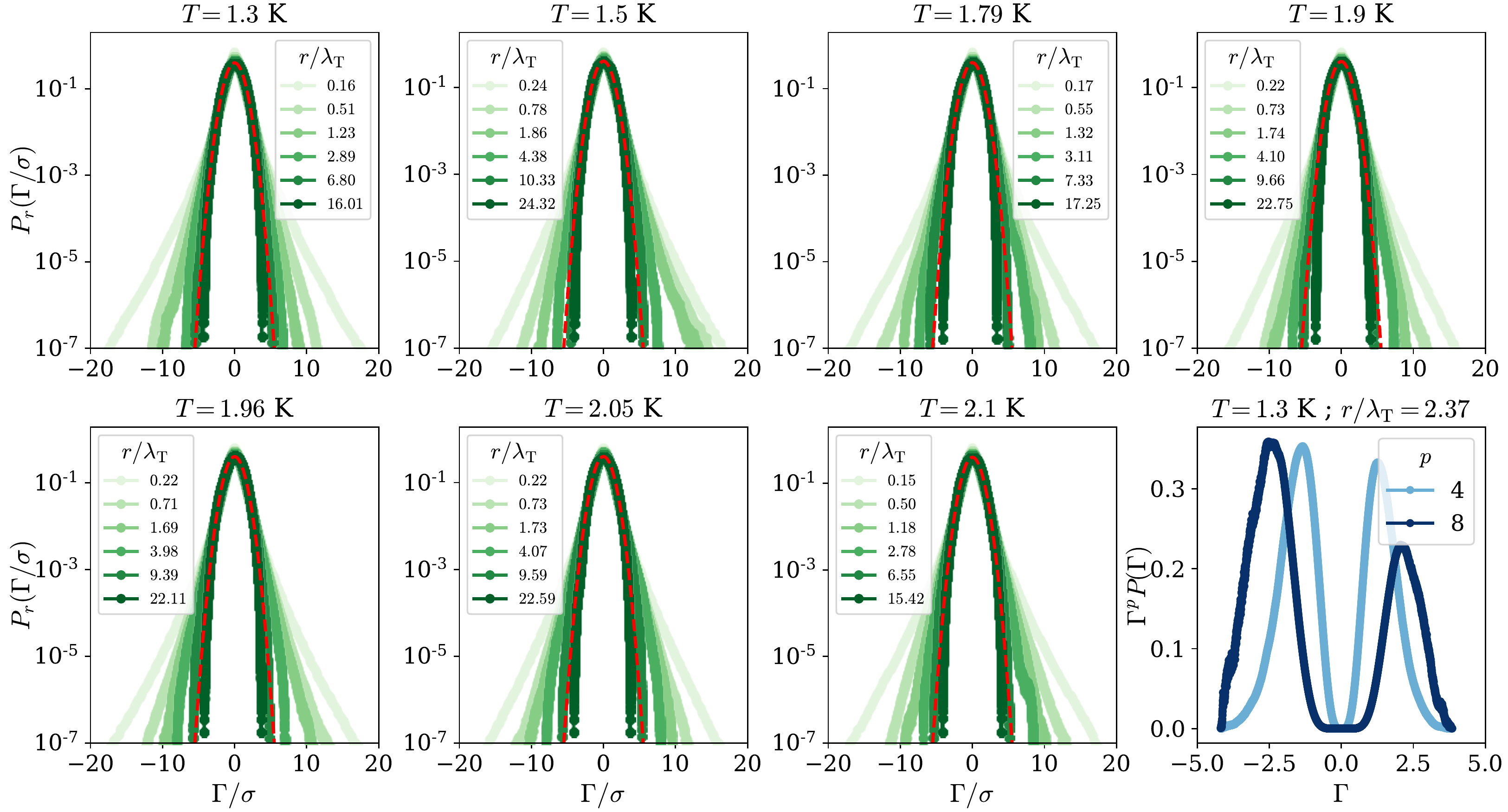}
  \caption[]{%
    Normalized PDFs of the velocity circulation $\Gamma$ for different loop sizes at different temperatures. In red dashed lines we show as reference a Normal distribution. The last panel shows the circulation integrands up to order eight for $T=1.3$ K and a length scale within the inertial range.
    }
    \label{fig:pdfs}
\end{figure}

The probability density functions (PDFs) of the total velocity circulation normalized by its standard deviation $\sigma = \mean{\Gamma^2}^{1/2}$ for different length scales are presented in Fig.~\ref{fig:pdfs}. 
Here, length scales are normalized by the lambda micro-scale $\taylor = \sqrt{5E / \Omega}$ with $E = \int v^2/2 \dd V$ the total energy of the system and $\Omega = \int |\vortvec|^2/2 \dd V $ the enstrophy. 
For all temperatures, the PDFs follow a qualitatively similar behavior as the one observed in the experiments discussed in Sec.~\ref{subsec:experimental_results} (Fig. \ref{fig:exp_integrands}), with heavy tails for small scales and approaching a Gaussian for large scales. The circulation integrands show a good convergence up to order eight.

\begin{figure}[t]
  \centering
  \includegraphics[width=1\textwidth]{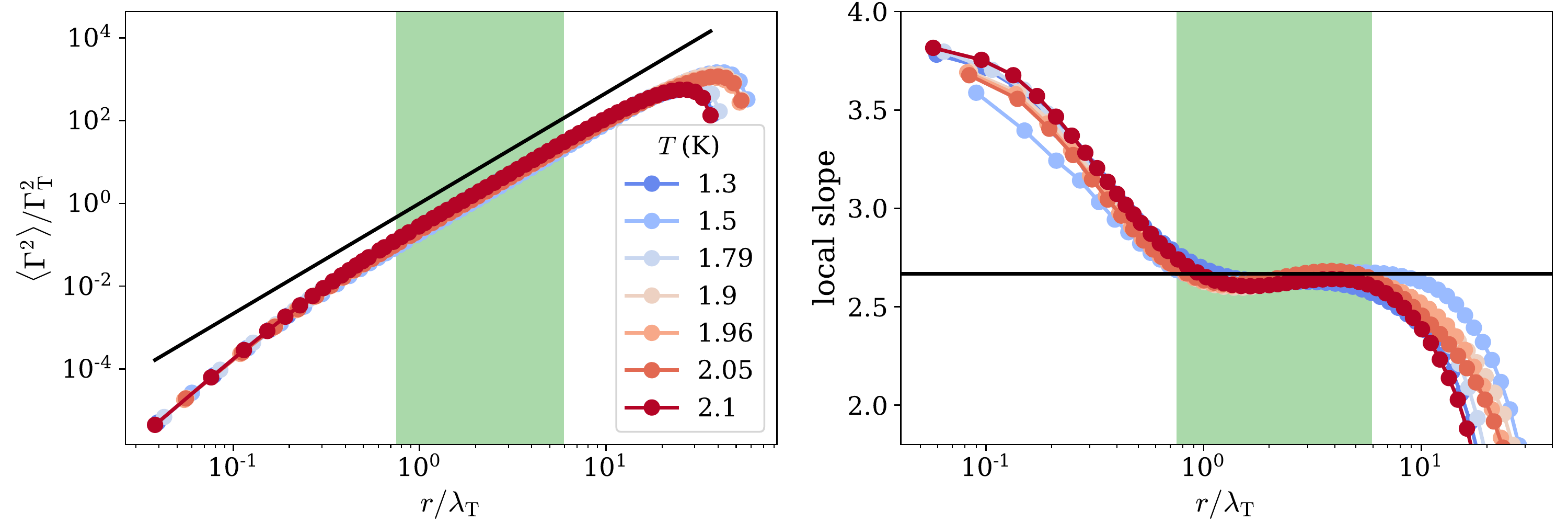}
  \caption[]{%
    Circulation variance of the total velocity $\vtotvec$ for different temperatures. Black solid lines correspond to Kolmogorov scaling. On the right, the local slope of the circulation variance, defined as the logarithmic derivative $\dd \log \mean{\Gamma^2} / \dd \log r$.
    }
    \label{fig:variance}
\end{figure}

\begin{figure}[t]
  \centering
  \includegraphics[width=1\textwidth]{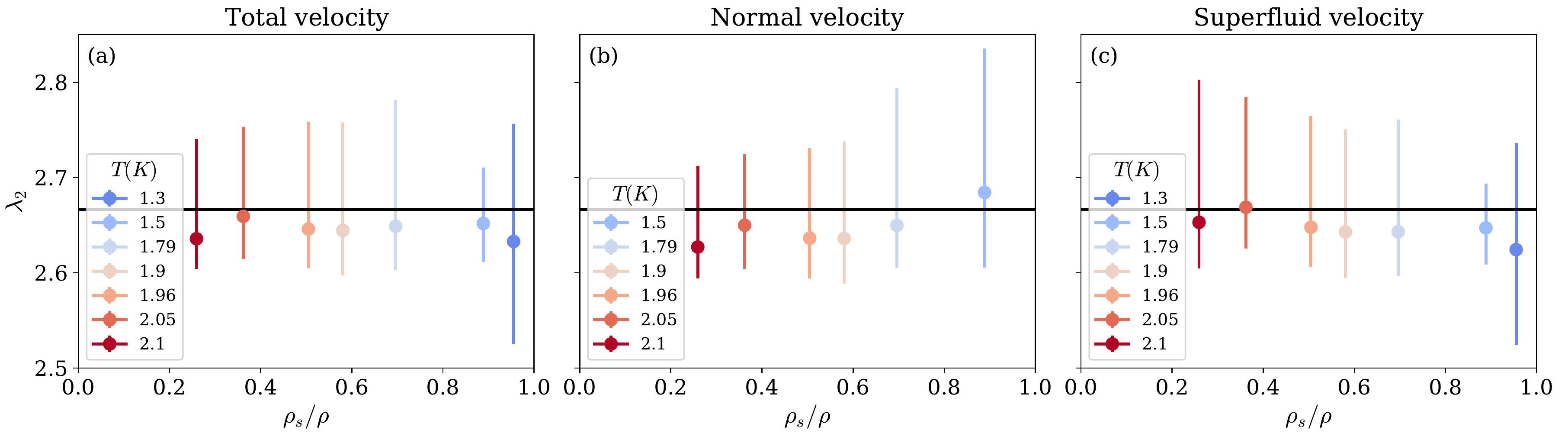}
  \caption[]{%
    Scaling exponents of the second order moment of the velocity circulation at different temperatures for (a) the total velocity, (b) the normal velocity and (c) the superfluid velocity fields. As a reference, the solid black line shows Kolmogorov scaling $\lambdaKolm_2 = 8/3$. Error bars indicate the maximum and minimum value of the local slope within the inertial range. The lowest temperature is removed from the middle panel due to the low mass density of the normal component.
    }
    \label{fig:lambda2}
\end{figure}

The circulation variance for different temperatures is shown in Fig. \ref{fig:variance}. The circulation is normalized by $\Gamma_{\mathrm{T}}^2 = (\taylor^4/3) \mean{|\vortvec|^2}$, that corresponds to the small-scale prediction \cite{Muller2021}. In this manner, when the normalized circulation variance is plotted as a function $r/\taylor$, the data collapses for all temperatures.
For each individual temperature, the inertial range extends to a full decade. 
The green region corresponds to the intersection of all inertial ranges, corresponding also to scales where $\mathcal{K}>0.95$. 
For all temperatures, the circulation variance follows a scaling close to Kolmogorov one $\mean{\Gamma^2} \sim r^{8/3}$. On the right panel we show that the local slope approaches a plateau of $8/3$ within the inertial range of scales. 
To analyse more in detail the temperature dependence of the system, we show the scaling exponents of the circulation variance as a function of the superfluid density $\rho_s / \rho$ for the different velocity components in Fig. \ref{fig:lambda2}. 
The error bars correspond to the maximum and minimum values of the local slope in the inertial range. 
The different velocity components display no significant difference between themselves, supporting the argument of velocity locking. 
Also, in all cases there is no apparent temperature dependence and the exponents approach to Kolmogorov $\lambda_2^{\mathrm{K41}} = 8/3$. The temperature $T=1.3$ K is removed from the normal velocity scaling due to the fact that the normal mass density is very small, displaying no clear scaling. 

\begin{figure}[t]
  \centering
  \includegraphics[width=.49\textwidth]{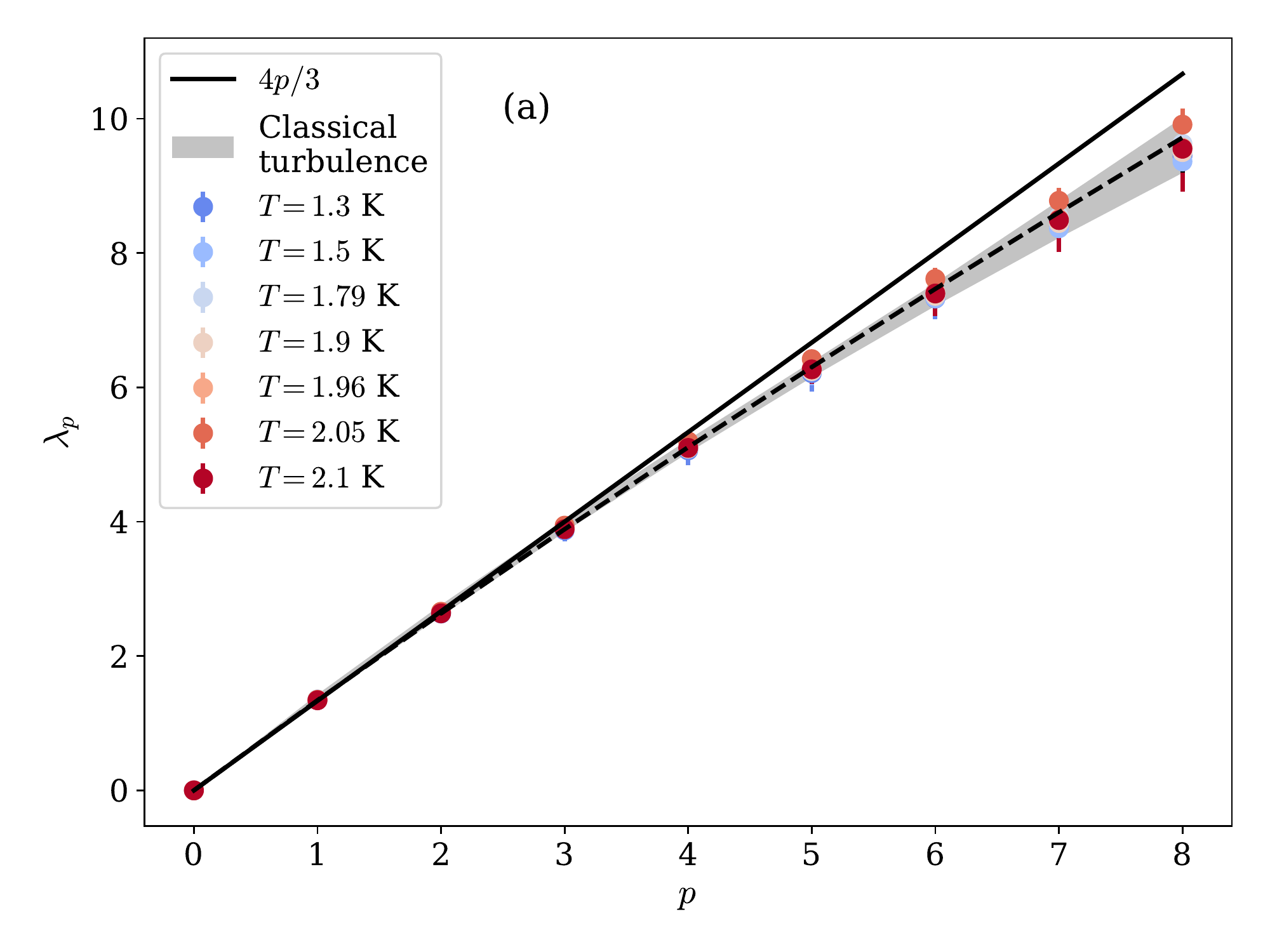}
  \includegraphics[width=.49\textwidth]{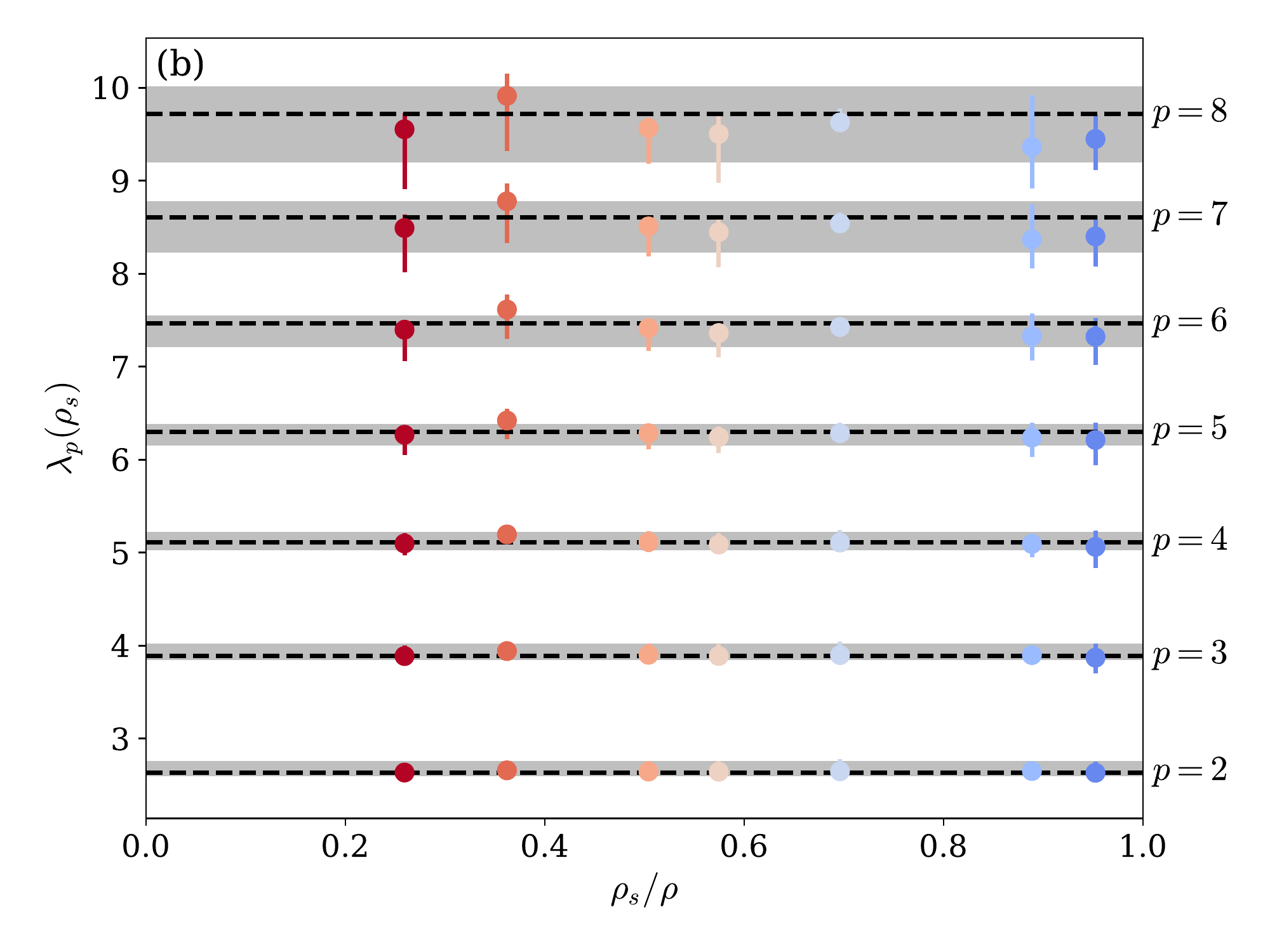}
  \caption[]{%
    Scaling exponent of the $p$-order moments of the velocity circulation at different temperatures. As a reference, the solid black line shows Kolmogorov scaling $\lambdaKolm_p = 4p/3$ and gray shaded area shows the scaling obtained from numerical simulations of the Navier--Stokes equations, with the black dashed line its mean value. 
    }
    \label{fig:exponents}
\end{figure}

For high-order moments, the scaling exponents of the system seem to follow the same behavior observed in numerical simulations of the Navier--Stokes and Gross--Pitaevskii equations \cite{Iyer2019, Muller2021}, the latter represented by the shaded area in Fig. \ref{fig:exponents} (a) which accounts for the error bars of data from \cite{Polanco2021a}. For $p \leq 3$, the scaling exponents of the velocity circulation follow Kolmogorov scaling $\lambdaKolm_p = 4p/3$, while for higher-order moments up to $p=8$ 
the scaling can be described by different multifractal models \cite{Iyer2019,Polanco2021a,Moriconi2021a}.
Figure \ref{fig:exponents} (b) shows the scaling exponents from $p=2$ to $p=8$ as a function of the superfluid density. Horizontal dashed lines correspond to the exponents obtained in classical turbulence, and the gray area including its error bars. Here, it is clear that there is no apparent temperature dependence on the circulation scaling even for high-order moments, following in all cases the same behavior as in classical turbulence.

\section{Discussion}
\label{sec:discussion}

In this work, we have addressed the scaling of circulation moments in superfluid helium at different temperatures. We have used superfluid grid turbulence experiments and numerical simulations of the HVBK model. We have compared the resulting circulation scaling exponents with those of Navier-Stokes (classical turbulence) and Gross-Pitaevskii (zero-temperature quantum turbulence) simulations from \cite{Polanco2021a}. 

\begin{figure}[t]
  \centering
  \includegraphics[width=.79\textwidth]{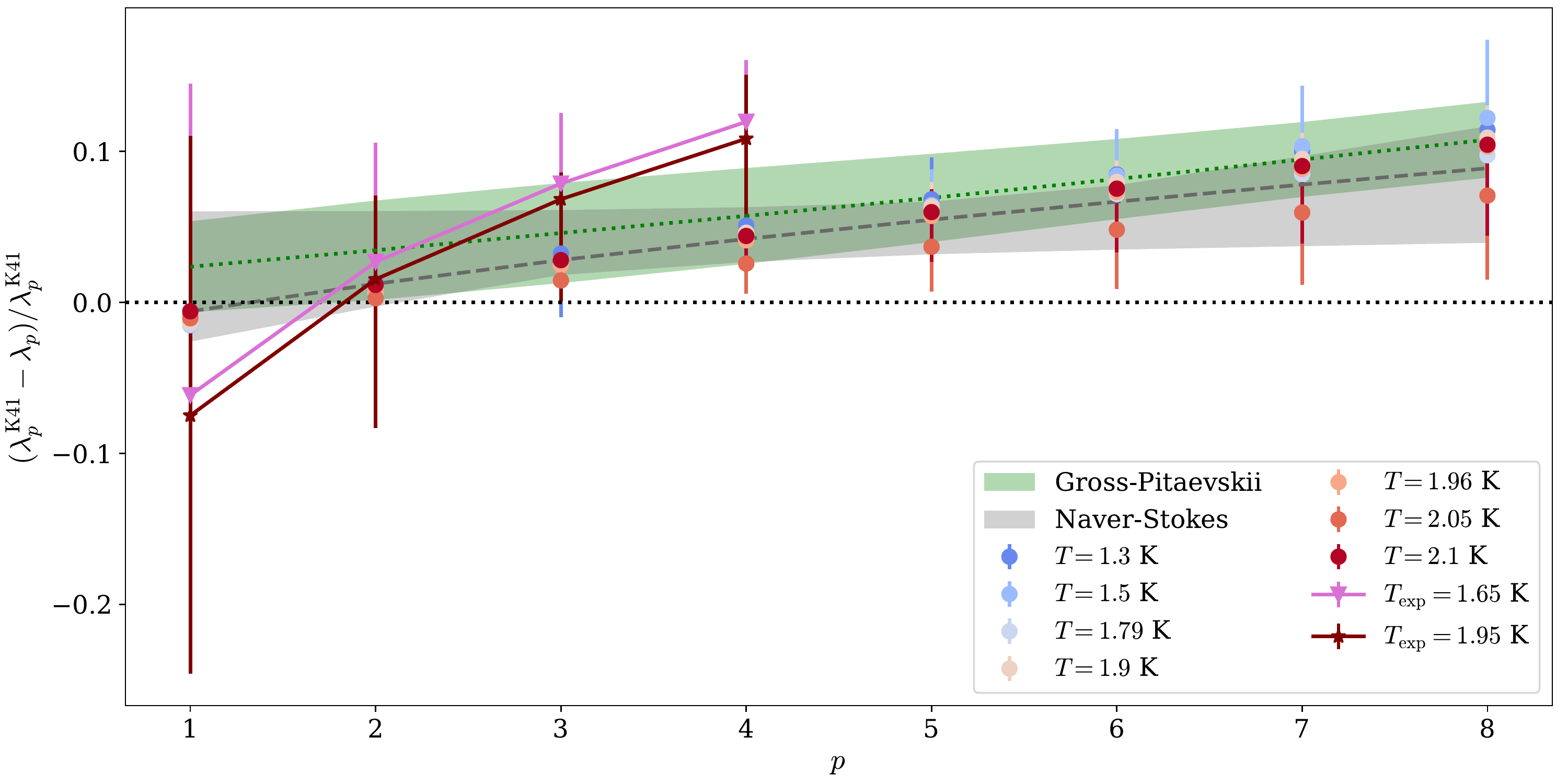}
  \caption[]{%
  Relative deviation of the circulation scaling exponents $\lambda_p$ with respect to K41 prediction $\lambdaKolm_p = 4p/3$. Data corresponds to superfluid grid turbulence experiments and numerical simulations of the HVBK, Gross-Pitaevskii and Navier-Stokes models. Grid superfluid experiments are realized at two different temperatures ($T_{\rm exp}=1.65$K and $T_{\rm exp}=1.95$K). Exponents obtained from HVBK data of current work are at temperatures $T=1.3$K - $2.1$K. Error bars are obtained by measuring the maximum and minimum of the local slope in the inertial range (see text). Classical and zero-temperature quantum turbulence exponents are taken from Navier-Stokes, and Gross-Pitaevskii simulations of reference \cite{Polanco2021a}. The green and gray areas show the error bars for those datasets, respectively.
    }
    \label{fig:exponentdeviations}
\end{figure}

We obtained the scaling exponents for experiments at temperatures $T=1.65$ K and $T=1.95$ K up to order four. Remarkably, we have observed a clear Kolmogorov scaling for the circulation variance, and there is no apparent temperature dependence within the error bars. For the HVBK numerical simulations, we have varied the temperature in the range $1.3 \leq T \leq 2.1$ K and observed that there is no clear temperature dependence neither on the intermittent behavior both for low and high-order moments of velocity circulation. Furthermore, experimental and HVBK data coincide, within error bars, with classical and low-temperature quantum turbulence simulations. Figure \ref{fig:exponentdeviations} presents the relative deviation $(\lambdaKolm_p - \lambda_p)/\lambdaKolm_p$ of the circulation exponents $\lambda_p$ with respect to the Kolmogorov scaling $\lambdaKolm_p$ for all available data.

Note that if one drops error bars, there is a slight systematic departure of the experimentally measured circulation exponents from those obtained using HVBK simulations. First, one could be tempted to claim that such a deviation origins from the HVBK description of superfluid helium which might fail to capture the whole physics of superfluids. Indeed, the HVBK model provides only a corse-grained description of superfluids and does not incorporate the dynamics of quantized vortices. Quantum vortices are related to singularities of the velocity field, which could impact high-order statistics. Whereas such singularities could affect velocity increments, they have no impact on circulation as it is perfectly well defined for quantum vortices (it is actually quantized); see \cite{Muller2021} for further discussion. Secondly, the available statistics used to compute high-order moments might not be enough to observe clean power laws in the inertial range, which could undoubtedly induce some errors. Finally, the circulation was computed using Eulerian fields constructed from Lagrangian particles. Several issues can arise from this method. For instance, a lack of particles in a given location of the flow could induce larger regions of constant velocity with abrupt jumps, affecting circulation values. Such regions are visible in the experimental Eulerian fields in Fig.~\ref{fig:exp_schematic} (b). Moreover, particles might be trapped by superfluid vortices \cite{Tang2021,Giuriato2019,Polanco2020}. In that case, they cannot be considered perfect tracers, which will affect the determination of the Eulerian fields for which this assumption is crucial. All those effects are difficult to quantify. On the other hand, the fact that the variance displays such a clear K41 scaling validates the current method and motivates its use for further studies. Whether the slight intermittency enhancement observed in experiments has an actual physical origin or arises from the construction of the Eulerian fields is an interesting question that should be addressed in the future.

\begin{acknowledgments}
  We are grateful to J.~I.~Polanco for fruitful scientific discussions. This work was supported by the Agence Nationale de la Recherche through the project GIANTE ANR-18-CE30-0020-01.
This work was granted access to the HPC resources of CINES, IDRIS and TGCC under the allocation 2019-A0072A11003 made by GENCI.
Computations were also carried out at the Mésocentre SIGAMM hosted at the Observatoire de la Côte d'Azur.
Y.T. and W.G. are supported by the National Science Foundation under Grant DMR-2100790 and the US Department of Energy under Grant DE-SC0020113.
They also acknowledge the support and resources provided by the National High Magnetic Field Laboratory at Florida State University, which is supported by the National Science Foundation Cooperative Agreement No. DMR-1644779 and the state of Florida.

\end{acknowledgments}

\bibliography{HVBK}

\begin{thebibliography}{48}%
\makeatletter
\providecommand \@ifxundefined [1]{%
 \@ifx{#1\undefined}
}%
\providecommand \@ifnum [1]{%
 \ifnum #1\expandafter \@firstoftwo
 \else \expandafter \@secondoftwo
 \fi
}%
\providecommand \@ifx [1]{%
 \ifx #1\expandafter \@firstoftwo
 \else \expandafter \@secondoftwo
 \fi
}%
\providecommand \natexlab [1]{#1}%
\providecommand \enquote  [1]{``#1''}%
\providecommand \bibnamefont  [1]{#1}%
\providecommand \bibfnamefont [1]{#1}%
\providecommand \citenamefont [1]{#1}%
\providecommand \href@noop [0]{\@secondoftwo}%
\providecommand \href [0]{\begingroup \@sanitize@url \@href}%
\providecommand \@href[1]{\@@startlink{#1}\@@href}%
\providecommand \@@href[1]{\endgroup#1\@@endlink}%
\providecommand \@sanitize@url [0]{\catcode `\\12\catcode `\$12\catcode
  `\&12\catcode `\#12\catcode `\^12\catcode `\_12\catcode `\%12\relax}%
\providecommand \@@startlink[1]{}%
\providecommand \@@endlink[0]{}%
\providecommand \url  [0]{\begingroup\@sanitize@url \@url }%
\providecommand \@url [1]{\endgroup\@href {#1}{\urlprefix }}%
\providecommand \urlprefix  [0]{URL }%
\providecommand \Eprint [0]{\href }%
\providecommand \doibase [0]{https://doi.org/}%
\providecommand \selectlanguage [0]{\@gobble}%
\providecommand \bibinfo  [0]{\@secondoftwo}%
\providecommand \bibfield  [0]{\@secondoftwo}%
\providecommand \translation [1]{[#1]}%
\providecommand \BibitemOpen [0]{}%
\providecommand \bibitemStop [0]{}%
\providecommand \bibitemNoStop [0]{.\EOS\space}%
\providecommand \EOS [0]{\spacefactor3000\relax}%
\providecommand \BibitemShut  [1]{\csname bibitem#1\endcsname}%
\let\auto@bib@innerbib\@empty
\bibitem [{\citenamefont {Davidson}(2013)}]{Davidson2013}%
  \BibitemOpen
  \bibfield  {author} {\bibinfo {author} {\bibfnamefont {P.~A.}\ \bibnamefont
  {Davidson}},\ }\href@noop {} {\emph {\bibinfo {title} {Turbulence in
  {{Rotating}}, {{Stratified}} and {{Electrically Conducting Fluids}}}}}\
  (\bibinfo  {publisher} {{Oxford university press}},\ \bibinfo {year}
  {2013})\BibitemShut {NoStop}%
\bibitem [{\citenamefont {Kolmogorov}(1941)}]{Kolmogorov1941a}%
  \BibitemOpen
  \bibfield  {author} {\bibinfo {author} {\bibfnamefont {A.~N.}\ \bibnamefont
  {Kolmogorov}},\ }\bibfield  {title} {\bibinfo {title} {The local structure of
  turbulence in incompressible viscous fluid for very large {{Reynolds}}
  numbers},\ }\href@noop {} {\bibfield  {journal} {\bibinfo  {journal} {Doklady
  Akademii Nauk}\ }\textbf {\bibinfo {volume} {30}},\ \bibinfo {pages} {301}
  (\bibinfo {year} {1941})}\BibitemShut {NoStop}%
\bibitem [{\citenamefont {Frisch}(1995)}]{Frisch1995}%
  \BibitemOpen
  \bibfield  {author} {\bibinfo {author} {\bibfnamefont {U.}~\bibnamefont
  {Frisch}},\ }\href {https://doi.org/10.1017/CBO9781139170666} {\emph
  {\bibinfo {title} {Turbulence: {{The Legacy}} of {{A}}.{{N}}.
  {{Kolmogorov}}}}},\ \bibinfo {edition} {1st}\ ed.\ (\bibinfo  {publisher}
  {{Cambridge University Press}},\ \bibinfo {year} {1995})\BibitemShut
  {NoStop}%
\bibitem [{\citenamefont {Kolmogorov}(1962)}]{Kolmogorov1962}%
  \BibitemOpen
  \bibfield  {author} {\bibinfo {author} {\bibfnamefont {A.~N.}\ \bibnamefont
  {Kolmogorov}},\ }\bibfield  {title} {\bibinfo {title} {A refinement of
  previous hypotheses concerning the local structure of turbulence in a viscous
  incompressible fluid at high {{Reynolds}} number},\ }\href
  {https://doi.org/10.1017/S0022112062000518} {\bibfield  {journal} {\bibinfo
  {journal} {Journal of Fluid Mechanics}\ }\textbf {\bibinfo {volume} {13}},\
  \bibinfo {pages} {82} (\bibinfo {year} {1962})}\BibitemShut {NoStop}%
\bibitem [{\citenamefont {Benzi}\ \emph {et~al.}(1984)\citenamefont {Benzi},
  \citenamefont {Paladin}, \citenamefont {Parisis},\ and\ \citenamefont
  {Vulpiani}}]{Benzi1984}%
  \BibitemOpen
  \bibfield  {author} {\bibinfo {author} {\bibfnamefont {R.}~\bibnamefont
  {Benzi}}, \bibinfo {author} {\bibfnamefont {G.}~\bibnamefont {Paladin}},
  \bibinfo {author} {\bibfnamefont {G.}~\bibnamefont {Parisis}},\ and\ \bibinfo
  {author} {\bibfnamefont {A.}~\bibnamefont {Vulpiani}},\ }\bibfield  {title}
  {\bibinfo {title} {On the multifractal nature of fully developed turbulence
  and chaotic systems},\ }\href@noop {} {\bibfield  {journal} {\bibinfo
  {journal} {Journal of Physics A: Mathematical and General}\ ,\ \bibinfo
  {pages} {12}} (\bibinfo {year} {1984})}\BibitemShut {NoStop}%
\bibitem [{\citenamefont {She}\ and\ \citenamefont
  {L{\'e}v{\^e}que}(1994)}]{She1994}%
  \BibitemOpen
  \bibfield  {author} {\bibinfo {author} {\bibfnamefont {Z.-S.}\ \bibnamefont
  {She}}\ and\ \bibinfo {author} {\bibfnamefont {E.}~\bibnamefont
  {L{\'e}v{\^e}que}},\ }\bibfield  {title} {\bibinfo {title} {Universal scaling
  laws in fully developed turbulence},\ }\href
  {https://doi.org/10.1103/PhysRevLett.72.336} {\bibfield  {journal} {\bibinfo
  {journal} {Physical Review Letters}\ }\textbf {\bibinfo {volume} {72}},\
  \bibinfo {pages} {336} (\bibinfo {year} {1994})}\BibitemShut {NoStop}%
\bibitem [{\citenamefont {Barenghi}\ \emph {et~al.}(2014)\citenamefont
  {Barenghi}, \citenamefont {Skrbek},\ and\ \citenamefont
  {Sreenivasan}}]{Barenghi2014}%
  \BibitemOpen
  \bibfield  {author} {\bibinfo {author} {\bibfnamefont {C.~F.}\ \bibnamefont
  {Barenghi}}, \bibinfo {author} {\bibfnamefont {L.}~\bibnamefont {Skrbek}},\
  and\ \bibinfo {author} {\bibfnamefont {K.~R.}\ \bibnamefont {Sreenivasan}},\
  }\bibfield  {title} {\bibinfo {title} {Introduction to quantum turbulence},\
  }\href {https://doi.org/10.1073/pnas.1400033111} {\bibfield  {journal}
  {\bibinfo  {journal} {Proceedings of the National Academy of Sciences of the
  United States of America}\ }\textbf {\bibinfo {volume} {111}},\ \bibinfo
  {pages} {4647} (\bibinfo {year} {2014})}\BibitemShut {NoStop}%
\bibitem [{\citenamefont {Pitaevskii}\ and\ \citenamefont
  {Stringari}(2016)}]{Pitaevskii2016}%
  \BibitemOpen
  \bibfield  {author} {\bibinfo {author} {\bibfnamefont {L.~P.}\ \bibnamefont
  {Pitaevskii}}\ and\ \bibinfo {author} {\bibfnamefont {S.}~\bibnamefont
  {Stringari}},\ }\href@noop {} {\emph {\bibinfo {title} {Bose-{{Einstein}}
  Condensation and Superfluidity}}},\ Vol.\ \bibinfo {volume} {164}\ (\bibinfo
  {publisher} {{Oxford University Press}},\ \bibinfo {year} {2016})\BibitemShut
  {NoStop}%
\bibitem [{\citenamefont {Donnelly}(1991)}]{Donnelly1991}%
  \BibitemOpen
  \bibfield  {author} {\bibinfo {author} {\bibfnamefont {R.~J.}\ \bibnamefont
  {Donnelly}},\ }\href@noop {} {\emph {\bibinfo {title} {Quantized {{Vortices}}
  in {{Helium II}}}}}\ (\bibinfo  {publisher} {{Cambridge University Press}},\
  \bibinfo {year} {1991})\BibitemShut {NoStop}%
\bibitem [{\citenamefont {Maurer}\ and\ \citenamefont
  {Tabeling}(1998)}]{Maurer1998}%
  \BibitemOpen
  \bibfield  {author} {\bibinfo {author} {\bibfnamefont {J.}~\bibnamefont
  {Maurer}}\ and\ \bibinfo {author} {\bibfnamefont {P.}~\bibnamefont
  {Tabeling}},\ }\bibfield  {title} {\bibinfo {title} {Local investigation of
  superfluid turbulence},\ }\href {https://doi.org/10.1209/epl/i1998-00314-9}
  {\bibfield  {journal} {\bibinfo  {journal} {Europhysics Letters (EPL)}\
  }\textbf {\bibinfo {volume} {43}},\ \bibinfo {pages} {29} (\bibinfo {year}
  {1998})}\BibitemShut {NoStop}%
\bibitem [{\citenamefont {Biferale}\ \emph {et~al.}(2019)\citenamefont
  {Biferale}, \citenamefont {Khomenko}, \citenamefont {L'vov}, \citenamefont
  {Pomyalov}, \citenamefont {Procaccia},\ and\ \citenamefont
  {Sahoo}}]{Biferale2019}%
  \BibitemOpen
  \bibfield  {author} {\bibinfo {author} {\bibfnamefont {L.}~\bibnamefont
  {Biferale}}, \bibinfo {author} {\bibfnamefont {D.}~\bibnamefont {Khomenko}},
  \bibinfo {author} {\bibfnamefont {V.}~\bibnamefont {L'vov}}, \bibinfo
  {author} {\bibfnamefont {A.}~\bibnamefont {Pomyalov}}, \bibinfo {author}
  {\bibfnamefont {I.}~\bibnamefont {Procaccia}},\ and\ \bibinfo {author}
  {\bibfnamefont {G.}~\bibnamefont {Sahoo}},\ }\bibfield  {title} {\bibinfo
  {title} {Superfluid {{Helium}} in {{Three-Dimensional Counterflow Differs
  Strongly}} from {{Classical Flows}}: {{Anisotropy}} on {{Small Scales}}},\
  }\href {https://doi.org/10.1103/PhysRevLett.122.144501} {\bibfield  {journal}
  {\bibinfo  {journal} {Physical Review Letters}\ }\textbf {\bibinfo {volume}
  {122}},\ \bibinfo {pages} {144501} (\bibinfo {year} {2019})}\BibitemShut
  {NoStop}%
\bibitem [{\citenamefont {Polanco}\ and\ \citenamefont
  {Krstulovic}(2020{\natexlab{a}})}]{Polanco2020a}%
  \BibitemOpen
  \bibfield  {author} {\bibinfo {author} {\bibfnamefont {J.~I.}\ \bibnamefont
  {Polanco}}\ and\ \bibinfo {author} {\bibfnamefont {G.}~\bibnamefont
  {Krstulovic}},\ }\bibfield  {title} {\bibinfo {title} {Counterflow-{{Induced
  Inverse Energy Cascade}} in {{Three-Dimensional Superfluid Turbulence}}},\
  }\href {https://doi.org/10.1103/PhysRevLett.125.254504} {\bibfield  {journal}
  {\bibinfo  {journal} {Physical Review Letters}\ }\textbf {\bibinfo {volume}
  {125}},\ \bibinfo {pages} {254504} (\bibinfo {year}
  {2020}{\natexlab{a}})}\BibitemShut {NoStop}%
\bibitem [{\citenamefont {Rusaouen}\ \emph {et~al.}(2017)\citenamefont
  {Rusaouen}, \citenamefont {Chabaud}, \citenamefont {Salort},\ and\
  \citenamefont {Roche}}]{Rusaouen2017}%
  \BibitemOpen
  \bibfield  {author} {\bibinfo {author} {\bibfnamefont {E.}~\bibnamefont
  {Rusaouen}}, \bibinfo {author} {\bibfnamefont {B.}~\bibnamefont {Chabaud}},
  \bibinfo {author} {\bibfnamefont {J.}~\bibnamefont {Salort}},\ and\ \bibinfo
  {author} {\bibfnamefont {P.-E.}\ \bibnamefont {Roche}},\ }\bibfield  {title}
  {\bibinfo {title} {Intermittency of quantum turbulence with superfluid
  fractions from 0\% to 96\%},\ }\href {https://doi.org/10.1063/1.4991558}
  {\bibfield  {journal} {\bibinfo  {journal} {Physics of Fluids}\ }\textbf
  {\bibinfo {volume} {29}},\ \bibinfo {pages} {105108} (\bibinfo {year}
  {2017})}\BibitemShut {NoStop}%
\bibitem [{\citenamefont {Varga}\ \emph {et~al.}(2018)\citenamefont {Varga},
  \citenamefont {Gao}, \citenamefont {Guo},\ and\ \citenamefont
  {Skrbek}}]{Varga2018}%
  \BibitemOpen
  \bibfield  {author} {\bibinfo {author} {\bibfnamefont {E.}~\bibnamefont
  {Varga}}, \bibinfo {author} {\bibfnamefont {J.}~\bibnamefont {Gao}}, \bibinfo
  {author} {\bibfnamefont {W.}~\bibnamefont {Guo}},\ and\ \bibinfo {author}
  {\bibfnamefont {L.}~\bibnamefont {Skrbek}},\ }\bibfield  {title} {\bibinfo
  {title} {Intermittency enhancement in quantum turbulence in superfluid {{He}}
  4},\ }\href {https://doi.org/10.1103/PhysRevFluids.3.094601} {\bibfield
  {journal} {\bibinfo  {journal} {Physical Review Fluids}\ }\textbf {\bibinfo
  {volume} {3}},\ \bibinfo {pages} {094601} (\bibinfo {year}
  {2018})}\BibitemShut {NoStop}%
\bibitem [{\citenamefont {Tang}\ \emph {et~al.}(2020)\citenamefont {Tang},
  \citenamefont {Bao}, \citenamefont {Kanai},\ and\ \citenamefont
  {Guo}}]{Tang2020}%
  \BibitemOpen
  \bibfield  {author} {\bibinfo {author} {\bibfnamefont {Y.}~\bibnamefont
  {Tang}}, \bibinfo {author} {\bibfnamefont {S.}~\bibnamefont {Bao}}, \bibinfo
  {author} {\bibfnamefont {T.}~\bibnamefont {Kanai}},\ and\ \bibinfo {author}
  {\bibfnamefont {W.}~\bibnamefont {Guo}},\ }\bibfield  {title} {\bibinfo
  {title} {Statistical properties of homogeneous and isotropic turbulence in
  {{He II}} measured via particle tracking velocimetry},\ }\href
  {https://doi.org/10.1103/PhysRevFluids.5.084602} {\bibfield  {journal}
  {\bibinfo  {journal} {Physical Review Fluids}\ }\textbf {\bibinfo {volume}
  {5}},\ \bibinfo {pages} {084602} (\bibinfo {year} {2020})}\BibitemShut
  {NoStop}%
\bibitem [{\citenamefont {Krstulovic}(2016)}]{Krstulovic2016}%
  \BibitemOpen
  \bibfield  {author} {\bibinfo {author} {\bibfnamefont {G.}~\bibnamefont
  {Krstulovic}},\ }\bibfield  {title} {\bibinfo {title} {Grid superfluid
  turbulence and intermittency at very low temperature},\ }\href
  {https://doi.org/10.1103/PhysRevE.93.063104} {\bibfield  {journal} {\bibinfo
  {journal} {Physical Review E}\ }\textbf {\bibinfo {volume} {93}},\ \bibinfo
  {pages} {063104} (\bibinfo {year} {2016})}\BibitemShut {NoStop}%
\bibitem [{\citenamefont {Biferale}\ \emph {et~al.}(2018)\citenamefont
  {Biferale}, \citenamefont {Khomenko}, \citenamefont {L'vov}, \citenamefont
  {Pomyalov}, \citenamefont {Procaccia},\ and\ \citenamefont
  {Sahoo}}]{Biferale2018a}%
  \BibitemOpen
  \bibfield  {author} {\bibinfo {author} {\bibfnamefont {L.}~\bibnamefont
  {Biferale}}, \bibinfo {author} {\bibfnamefont {D.}~\bibnamefont {Khomenko}},
  \bibinfo {author} {\bibfnamefont {V.}~\bibnamefont {L'vov}}, \bibinfo
  {author} {\bibfnamefont {A.}~\bibnamefont {Pomyalov}}, \bibinfo {author}
  {\bibfnamefont {I.}~\bibnamefont {Procaccia}},\ and\ \bibinfo {author}
  {\bibfnamefont {G.}~\bibnamefont {Sahoo}},\ }\bibfield  {title} {\bibinfo
  {title} {Turbulent statistics and intermittency enhancement in coflowing
  superfluid {{He}} 4},\ }\href
  {https://doi.org/10.1103/PhysRevFluids.3.024605} {\bibfield  {journal}
  {\bibinfo  {journal} {Physical Review Fluids}\ }\textbf {\bibinfo {volume}
  {3}},\ \bibinfo {pages} {024605} (\bibinfo {year} {2018})}\BibitemShut
  {NoStop}%
\bibitem [{\citenamefont {Bou{\'e}}\ \emph {et~al.}(2013)\citenamefont
  {Bou{\'e}}, \citenamefont {L'vov}, \citenamefont {Pomyalov},\ and\
  \citenamefont {Procaccia}}]{Boue2013}%
  \BibitemOpen
  \bibfield  {author} {\bibinfo {author} {\bibfnamefont {L.}~\bibnamefont
  {Bou{\'e}}}, \bibinfo {author} {\bibfnamefont {V.}~\bibnamefont {L'vov}},
  \bibinfo {author} {\bibfnamefont {A.}~\bibnamefont {Pomyalov}},\ and\
  \bibinfo {author} {\bibfnamefont {I.}~\bibnamefont {Procaccia}},\ }\bibfield
  {title} {\bibinfo {title} {Enhancement of {{Intermittency}} in {{Superfluid
  Turbulence}}},\ }\href {https://doi.org/10.1103/PhysRevLett.110.014502}
  {\bibfield  {journal} {\bibinfo  {journal} {Physical Review Letters}\
  }\textbf {\bibinfo {volume} {110}},\ \bibinfo {pages} {014502} (\bibinfo
  {year} {2013})}\BibitemShut {NoStop}%
\bibitem [{\citenamefont {Shukla}\ and\ \citenamefont
  {Pandit}(2016)}]{Shukla2016}%
  \BibitemOpen
  \bibfield  {author} {\bibinfo {author} {\bibfnamefont {V.}~\bibnamefont
  {Shukla}}\ and\ \bibinfo {author} {\bibfnamefont {R.}~\bibnamefont
  {Pandit}},\ }\bibfield  {title} {\bibinfo {title} {Multiscaling in superfluid
  turbulence: {{A}} shell-model study},\ }\href
  {https://doi.org/10.1103/PhysRevE.94.043101} {\bibfield  {journal} {\bibinfo
  {journal} {Physical Review E}\ }\textbf {\bibinfo {volume} {94}},\ \bibinfo
  {pages} {043101} (\bibinfo {year} {2016})}\BibitemShut {NoStop}%
\bibitem [{\citenamefont {Migdal}(2020)}]{Migdal2020}%
  \BibitemOpen
  \bibfield  {author} {\bibinfo {author} {\bibfnamefont {A.}~\bibnamefont
  {Migdal}},\ }\bibfield  {title} {\bibinfo {title} {Clebsch confinement and
  instantons in turbulence},\ }\href
  {https://doi.org/10.1142/S0217751X20300185} {\bibfield  {journal} {\bibinfo
  {journal} {International Journal of Modern Physics A}\ }\textbf {\bibinfo
  {volume} {35}},\ \bibinfo {pages} {2030018} (\bibinfo {year}
  {2020})}\BibitemShut {NoStop}%
\bibitem [{\citenamefont {Sreenivasan}\ \emph {et~al.}(1995)\citenamefont
  {Sreenivasan}, \citenamefont {Juneja},\ and\ \citenamefont
  {Suri}}]{Sreenivasan1995}%
  \BibitemOpen
  \bibfield  {author} {\bibinfo {author} {\bibfnamefont {K.~R.}\ \bibnamefont
  {Sreenivasan}}, \bibinfo {author} {\bibfnamefont {A.}~\bibnamefont
  {Juneja}},\ and\ \bibinfo {author} {\bibfnamefont {A.~K.}\ \bibnamefont
  {Suri}},\ }\bibfield  {title} {\bibinfo {title} {Scaling {{Properties}} of
  {{Circulation}} in {{Moderate-Reynolds-Number Turbulent Wakes}}},\ }\href
  {https://doi.org/10.1103/PhysRevLett.75.433} {\bibfield  {journal} {\bibinfo
  {journal} {Physical Review Letters}\ }\textbf {\bibinfo {volume} {75}},\
  \bibinfo {pages} {433} (\bibinfo {year} {1995})}\BibitemShut {NoStop}%
\bibitem [{\citenamefont {Cao}\ \emph {et~al.}(1996)\citenamefont {Cao},
  \citenamefont {Chen},\ and\ \citenamefont {Sreenivasan}}]{Cao1996}%
  \BibitemOpen
  \bibfield  {author} {\bibinfo {author} {\bibfnamefont {N.}~\bibnamefont
  {Cao}}, \bibinfo {author} {\bibfnamefont {S.}~\bibnamefont {Chen}},\ and\
  \bibinfo {author} {\bibfnamefont {K.~R.}\ \bibnamefont {Sreenivasan}},\
  }\bibfield  {title} {\bibinfo {title} {Properties of {{Velocity Circulation}}
  in {{Three-Dimensional Turbulence}}},\ }\href
  {https://doi.org/10.1103/PhysRevLett.76.616} {\bibfield  {journal} {\bibinfo
  {journal} {Physical Review Letters}\ }\textbf {\bibinfo {volume} {76}},\
  \bibinfo {pages} {616} (\bibinfo {year} {1996})}\BibitemShut {NoStop}%
\bibitem [{\citenamefont {Benzi}\ \emph {et~al.}(1997)\citenamefont {Benzi},
  \citenamefont {Biferale}, \citenamefont {Struglia},\ and\ \citenamefont
  {Tripiccione}}]{Benzi1997}%
  \BibitemOpen
  \bibfield  {author} {\bibinfo {author} {\bibfnamefont {R.}~\bibnamefont
  {Benzi}}, \bibinfo {author} {\bibfnamefont {L.}~\bibnamefont {Biferale}},
  \bibinfo {author} {\bibfnamefont {M.~V.}\ \bibnamefont {Struglia}},\ and\
  \bibinfo {author} {\bibfnamefont {R.}~\bibnamefont {Tripiccione}},\
  }\bibfield  {title} {\bibinfo {title} {Self-scaling properties of velocity
  circulation in shear flows},\ }\href
  {https://doi.org/10.1103/PhysRevE.55.3739} {\bibfield  {journal} {\bibinfo
  {journal} {Physical Review E}\ }\textbf {\bibinfo {volume} {55}},\ \bibinfo
  {pages} {3739} (\bibinfo {year} {1997})}\BibitemShut {NoStop}%
\bibitem [{\citenamefont {Iyer}\ \emph {et~al.}(2021)\citenamefont {Iyer},
  \citenamefont {Bharadwaj},\ and\ \citenamefont {Sreenivasan}}]{Iyer2021}%
  \BibitemOpen
  \bibfield  {author} {\bibinfo {author} {\bibfnamefont {K.~P.}\ \bibnamefont
  {Iyer}}, \bibinfo {author} {\bibfnamefont {S.~S.}\ \bibnamefont
  {Bharadwaj}},\ and\ \bibinfo {author} {\bibfnamefont {K.~R.}\ \bibnamefont
  {Sreenivasan}},\ }\bibfield  {title} {\bibinfo {title} {The area rule for
  circulation in three-dimensional turbulence},\ }\href
  {https://doi.org/10.1073/pnas.2114679118} {\bibfield  {journal} {\bibinfo
  {journal} {Proceedings of the National Academy of Sciences}\ }\textbf
  {\bibinfo {volume} {118}},\ \bibinfo {pages} {e2114679118} (\bibinfo {year}
  {2021})}\BibitemShut {NoStop}%
\bibitem [{\citenamefont {Zhou}\ \emph {et~al.}(2008)\citenamefont {Zhou},
  \citenamefont {Sun},\ and\ \citenamefont {Xia}}]{Zhou2008}%
  \BibitemOpen
  \bibfield  {author} {\bibinfo {author} {\bibfnamefont {Q.}~\bibnamefont
  {Zhou}}, \bibinfo {author} {\bibfnamefont {C.}~\bibnamefont {Sun}},\ and\
  \bibinfo {author} {\bibfnamefont {K.-Q.}\ \bibnamefont {Xia}},\ }\bibfield
  {title} {\bibinfo {title} {Experimental investigation of homogeneity,
  isotropy, and circulation of the velocity field in buoyancy-driven
  turbulence},\ }\href {https://doi.org/10.1017/S0022112008000189} {\bibfield
  {journal} {\bibinfo  {journal} {Journal of Fluid Mechanics}\ }\textbf
  {\bibinfo {volume} {598}},\ \bibinfo {pages} {361} (\bibinfo {year}
  {2008})}\BibitemShut {NoStop}%
\bibitem [{\citenamefont {Iyer}\ \emph {et~al.}(2019)\citenamefont {Iyer},
  \citenamefont {Sreenivasan},\ and\ \citenamefont {Yeung}}]{Iyer2019}%
  \BibitemOpen
  \bibfield  {author} {\bibinfo {author} {\bibfnamefont {K.~P.}\ \bibnamefont
  {Iyer}}, \bibinfo {author} {\bibfnamefont {K.~R.}\ \bibnamefont
  {Sreenivasan}},\ and\ \bibinfo {author} {\bibfnamefont {P.~K.}\ \bibnamefont
  {Yeung}},\ }\bibfield  {title} {\bibinfo {title} {Circulation in {{High
  Reynolds Number Isotropic Turbulence}} is a {{Bifractal}}},\ }\href
  {https://doi.org/10.1103/PhysRevX.9.041006} {\bibfield  {journal} {\bibinfo
  {journal} {Physical Review X}\ }\textbf {\bibinfo {volume} {9}},\ \bibinfo
  {pages} {041006} (\bibinfo {year} {2019})}\BibitemShut {NoStop}%
\bibitem [{\citenamefont {M{\"u}ller}\ \emph {et~al.}(2021)\citenamefont
  {M{\"u}ller}, \citenamefont {Polanco},\ and\ \citenamefont
  {Krstulovic}}]{Muller2021}%
  \BibitemOpen
  \bibfield  {author} {\bibinfo {author} {\bibfnamefont {N.~P.}\ \bibnamefont
  {M{\"u}ller}}, \bibinfo {author} {\bibfnamefont {J.~I.}\ \bibnamefont
  {Polanco}},\ and\ \bibinfo {author} {\bibfnamefont {G.}~\bibnamefont
  {Krstulovic}},\ }\bibfield  {title} {\bibinfo {title} {Intermittency of
  {{Velocity Circulation}} in {{Quantum Turbulence}}},\ }\href
  {https://doi.org/10.1103/PhysRevX.11.011053} {\bibfield  {journal} {\bibinfo
  {journal} {Physical Review X}\ }\textbf {\bibinfo {volume} {11}},\ \bibinfo
  {pages} {011053} (\bibinfo {year} {2021})}\BibitemShut {NoStop}%
\bibitem [{\citenamefont {Polanco}\ \emph
  {et~al.}(2021{\natexlab{a}})\citenamefont {Polanco}, \citenamefont
  {M{\"u}ller},\ and\ \citenamefont {Krstulovic}}]{Polanco2021a}%
  \BibitemOpen
  \bibfield  {author} {\bibinfo {author} {\bibfnamefont {J.~I.}\ \bibnamefont
  {Polanco}}, \bibinfo {author} {\bibfnamefont {N.~P.}\ \bibnamefont
  {M{\"u}ller}},\ and\ \bibinfo {author} {\bibfnamefont {G.}~\bibnamefont
  {Krstulovic}},\ }\bibfield  {title} {\bibinfo {title} {Vortex clustering,
  polarisation and circulation intermittency in classical and quantum
  turbulence},\ }\href {https://doi.org/10.1038/s41467-021-27382-6} {\bibfield
  {journal} {\bibinfo  {journal} {Nature Communications}\ }\textbf {\bibinfo
  {volume} {12}},\ \bibinfo {pages} {7090} (\bibinfo {year}
  {2021}{\natexlab{a}})}\BibitemShut {NoStop}%
\bibitem [{\citenamefont {Moriconi}(2021)}]{Moriconi2021a}%
  \BibitemOpen
  \bibfield  {author} {\bibinfo {author} {\bibfnamefont {L.}~\bibnamefont
  {Moriconi}},\ }\bibfield  {title} {\bibinfo {title} {Multifractality breaking
  from bounded random measures},\ }\href
  {https://doi.org/10.1103/PhysRevE.103.062137} {\bibfield  {journal} {\bibinfo
   {journal} {Physical Review E}\ }\textbf {\bibinfo {volume} {103}},\ \bibinfo
  {pages} {062137} (\bibinfo {year} {2021})}\BibitemShut {NoStop}%
\bibitem [{\citenamefont {Mastracci}\ and\ \citenamefont
  {Guo}(2018{\natexlab{a}})}]{Mastracci2018}%
  \BibitemOpen
  \bibfield  {author} {\bibinfo {author} {\bibfnamefont {B.}~\bibnamefont
  {Mastracci}}\ and\ \bibinfo {author} {\bibfnamefont {W.}~\bibnamefont
  {Guo}},\ }\bibfield  {title} {\bibinfo {title} {An apparatus for generation
  and quantitative measurement of homogeneous isotropic turbulence in {{He}}
  {\textsc{ii}}},\ }\href {https://doi.org/10.1063/1.4997735} {\bibfield
  {journal} {\bibinfo  {journal} {Review of Scientific Instruments}\ }\textbf
  {\bibinfo {volume} {89}},\ \bibinfo {pages} {015107} (\bibinfo {year}
  {2018}{\natexlab{a}})}\BibitemShut {NoStop}%
\bibitem [{\citenamefont {Mastracci}\ and\ \citenamefont
  {Guo}(2018{\natexlab{b}})}]{Mastracci2018a}%
  \BibitemOpen
  \bibfield  {author} {\bibinfo {author} {\bibfnamefont {B.}~\bibnamefont
  {Mastracci}}\ and\ \bibinfo {author} {\bibfnamefont {W.}~\bibnamefont
  {Guo}},\ }\bibfield  {title} {\bibinfo {title} {Exploration of thermal
  counterflow in {{He II}} using particle tracking velocimetry},\ }\href
  {https://doi.org/10.1103/PhysRevFluids.3.063304} {\bibfield  {journal}
  {\bibinfo  {journal} {Physical Review Fluids}\ }\textbf {\bibinfo {volume}
  {3}},\ \bibinfo {pages} {063304} (\bibinfo {year}
  {2018}{\natexlab{b}})}\BibitemShut {NoStop}%
\bibitem [{\citenamefont {Polanco}\ and\ \citenamefont
  {Krstulovic}(2020{\natexlab{b}})}]{Polanco2020}%
  \BibitemOpen
  \bibfield  {author} {\bibinfo {author} {\bibfnamefont {J.~I.}\ \bibnamefont
  {Polanco}}\ and\ \bibinfo {author} {\bibfnamefont {G.}~\bibnamefont
  {Krstulovic}},\ }\bibfield  {title} {\bibinfo {title} {Inhomogeneous
  distribution of particles in coflow and counterflow quantum turbulence},\
  }\href {https://doi.org/10.1103/PhysRevFluids.5.032601} {\bibfield  {journal}
  {\bibinfo  {journal} {Physical Review Fluids}\ }\textbf {\bibinfo {volume}
  {5}},\ \bibinfo {pages} {032601(R)} (\bibinfo {year}
  {2020}{\natexlab{b}})}\BibitemShut {NoStop}%
\bibitem [{\citenamefont {Mastracci}\ and\ \citenamefont
  {Guo}(2019)}]{Mastracci2019}%
  \BibitemOpen
  \bibfield  {author} {\bibinfo {author} {\bibfnamefont {B.}~\bibnamefont
  {Mastracci}}\ and\ \bibinfo {author} {\bibfnamefont {W.}~\bibnamefont
  {Guo}},\ }\bibfield  {title} {\bibinfo {title} {Characterizing vortex tangle
  properties in steady-state {{He II}} counterflow using particle tracking
  velocimetry},\ }\href {https://doi.org/10.1103/PhysRevFluids.4.023301}
  {\bibfield  {journal} {\bibinfo  {journal} {Physical Review Fluids}\ }\textbf
  {\bibinfo {volume} {4}},\ \bibinfo {pages} {023301} (\bibinfo {year}
  {2019})}\BibitemShut {NoStop}%
\bibitem [{\citenamefont {Tang}\ \emph
  {et~al.}(2021{\natexlab{a}})\citenamefont {Tang}, \citenamefont {Bao},\ and\
  \citenamefont {Guo}}]{Tang2021}%
  \BibitemOpen
  \bibfield  {author} {\bibinfo {author} {\bibfnamefont {Y.}~\bibnamefont
  {Tang}}, \bibinfo {author} {\bibfnamefont {S.}~\bibnamefont {Bao}},\ and\
  \bibinfo {author} {\bibfnamefont {W.}~\bibnamefont {Guo}},\ }\bibfield
  {title} {\bibinfo {title} {Superdiffusion of quantized vortices uncovering
  scaling laws in quantum turbulence},\ }\href
  {https://doi.org/10.1073/pnas.2021957118} {\bibfield  {journal} {\bibinfo
  {journal} {Proceedings of the National Academy of Sciences}\ }\textbf
  {\bibinfo {volume} {118}},\ \bibinfo {pages} {e2021957118} (\bibinfo {year}
  {2021}{\natexlab{a}})}\BibitemShut {NoStop}%
\bibitem [{\citenamefont {Giuriato}\ and\ \citenamefont
  {Krstulovic}(2019)}]{Giuriato2019}%
  \BibitemOpen
  \bibfield  {author} {\bibinfo {author} {\bibfnamefont {U.}~\bibnamefont
  {Giuriato}}\ and\ \bibinfo {author} {\bibfnamefont {G.}~\bibnamefont
  {Krstulovic}},\ }\bibfield  {title} {\bibinfo {title} {Interaction between
  active particles and quantum vortices leading to {{Kelvin}} wave
  generation},\ }\href {https://doi.org/10.1038/s41598-019-39877-w} {\bibfield
  {journal} {\bibinfo  {journal} {Scientific Reports}\ }\textbf {\bibinfo
  {volume} {9}},\ \bibinfo {pages} {4839} (\bibinfo {year} {2019})}\BibitemShut
  {NoStop}%
\bibitem [{\citenamefont {Giuriato}\ and\ \citenamefont
  {Krstulovic}(2020)}]{Giuriato2020}%
  \BibitemOpen
  \bibfield  {author} {\bibinfo {author} {\bibfnamefont {U.}~\bibnamefont
  {Giuriato}}\ and\ \bibinfo {author} {\bibfnamefont {G.}~\bibnamefont
  {Krstulovic}},\ }\bibfield  {title} {\bibinfo {title} {Active and finite-size
  particles in decaying quantum turbulence at low temperature},\ }\href
  {https://doi.org/10.1103/PhysRevFluids.5.054608} {\bibfield  {journal}
  {\bibinfo  {journal} {Physical Review Fluids}\ }\textbf {\bibinfo {volume}
  {5}},\ \bibinfo {pages} {054608} (\bibinfo {year} {2020})}\BibitemShut
  {NoStop}%
\bibitem [{\citenamefont {Tang}\ \emph
  {et~al.}(2021{\natexlab{b}})\citenamefont {Tang}, \citenamefont {Guo},
  \citenamefont {L'vov},\ and\ \citenamefont {Pomyalov}}]{Tang2021a}%
  \BibitemOpen
  \bibfield  {author} {\bibinfo {author} {\bibfnamefont {Y.}~\bibnamefont
  {Tang}}, \bibinfo {author} {\bibfnamefont {W.}~\bibnamefont {Guo}}, \bibinfo
  {author} {\bibfnamefont {V.~S.}\ \bibnamefont {L'vov}},\ and\ \bibinfo
  {author} {\bibfnamefont {A.}~\bibnamefont {Pomyalov}},\ }\bibfield  {title}
  {\bibinfo {title} {Eulerian and {{Lagrangian}} second-order statistics of
  superfluid {{He}} 4 grid turbulence},\ }\href
  {https://doi.org/10.1103/PhysRevB.103.144506} {\bibfield  {journal} {\bibinfo
   {journal} {Physical Review B}\ }\textbf {\bibinfo {volume} {103}},\ \bibinfo
  {pages} {144506} (\bibinfo {year} {2021}{\natexlab{b}})}\BibitemShut
  {NoStop}%
\bibitem [{\citenamefont {Biferale}\ \emph {et~al.}(2017)\citenamefont
  {Biferale}, \citenamefont {Khomenko}, \citenamefont {L'vov}, \citenamefont
  {Pomyalov}, \citenamefont {Procaccia},\ and\ \citenamefont
  {Sahoo}}]{Biferale2017}%
  \BibitemOpen
  \bibfield  {author} {\bibinfo {author} {\bibfnamefont {L.}~\bibnamefont
  {Biferale}}, \bibinfo {author} {\bibfnamefont {D.}~\bibnamefont {Khomenko}},
  \bibinfo {author} {\bibfnamefont {V.}~\bibnamefont {L'vov}}, \bibinfo
  {author} {\bibfnamefont {A.}~\bibnamefont {Pomyalov}}, \bibinfo {author}
  {\bibfnamefont {I.}~\bibnamefont {Procaccia}},\ and\ \bibinfo {author}
  {\bibfnamefont {G.}~\bibnamefont {Sahoo}},\ }\bibfield  {title} {\bibinfo
  {title} {Local and nonlocal energy spectra of superfluid {{He}} 3
  turbulence},\ }\href {https://doi.org/10.1103/PhysRevB.95.184510} {\bibfield
  {journal} {\bibinfo  {journal} {Physical Review B}\ }\textbf {\bibinfo
  {volume} {95}},\ \bibinfo {pages} {184510} (\bibinfo {year}
  {2017})}\BibitemShut {NoStop}%
\bibitem [{\citenamefont {Koplik}\ and\ \citenamefont
  {Levine}(1993)}]{Koplik1993}%
  \BibitemOpen
  \bibfield  {author} {\bibinfo {author} {\bibfnamefont {J.}~\bibnamefont
  {Koplik}}\ and\ \bibinfo {author} {\bibfnamefont {H.}~\bibnamefont
  {Levine}},\ }\bibfield  {title} {\bibinfo {title} {Vortex reconnection in
  superfluid helium},\ }\href {https://doi.org/10.1103/PhysRevLett.71.1375}
  {\bibfield  {journal} {\bibinfo  {journal} {Physical Review Letters}\
  }\textbf {\bibinfo {volume} {71}},\ \bibinfo {pages} {1375} (\bibinfo {year}
  {1993})}\BibitemShut {NoStop}%
\bibitem [{\citenamefont {Bewley}\ \emph {et~al.}(2008)\citenamefont {Bewley},
  \citenamefont {Paoletti}, \citenamefont {Sreenivasan},\ and\ \citenamefont
  {Lathrop}}]{Bewley2008}%
  \BibitemOpen
  \bibfield  {author} {\bibinfo {author} {\bibfnamefont {G.~P.}\ \bibnamefont
  {Bewley}}, \bibinfo {author} {\bibfnamefont {M.~S.}\ \bibnamefont
  {Paoletti}}, \bibinfo {author} {\bibfnamefont {K.~R.}\ \bibnamefont
  {Sreenivasan}},\ and\ \bibinfo {author} {\bibfnamefont {D.~P.}\ \bibnamefont
  {Lathrop}},\ }\bibfield  {title} {\bibinfo {title} {Characterization of
  reconnecting vortices in superfluid helium},\ }\href
  {https://doi.org/10.1073/pnas.0806002105} {\bibfield  {journal} {\bibinfo
  {journal} {Proceedings of the National Academy of Sciences of the United
  States of America}\ }\textbf {\bibinfo {volume} {105}},\ \bibinfo {pages}
  {13707} (\bibinfo {year} {2008})}\BibitemShut {NoStop}%
\bibitem [{\citenamefont {Villois}\ \emph {et~al.}(2020)\citenamefont
  {Villois}, \citenamefont {Proment},\ and\ \citenamefont
  {Krstulovic}}]{Villois2020}%
  \BibitemOpen
  \bibfield  {author} {\bibinfo {author} {\bibfnamefont {A.}~\bibnamefont
  {Villois}}, \bibinfo {author} {\bibfnamefont {D.}~\bibnamefont {Proment}},\
  and\ \bibinfo {author} {\bibfnamefont {G.}~\bibnamefont {Krstulovic}},\
  }\bibfield  {title} {\bibinfo {title} {Irreversible {{Dynamics}} of {{Vortex
  Reconnections}} in {{Quantum Fluids}}},\ }\href
  {https://doi.org/10.1103/PhysRevLett.125.164501} {\bibfield  {journal}
  {\bibinfo  {journal} {Physical Review Letters}\ }\textbf {\bibinfo {volume}
  {125}},\ \bibinfo {pages} {164501} (\bibinfo {year} {2020})}\BibitemShut
  {NoStop}%
\bibitem [{\citenamefont {Krstulovic}(2012)}]{Krstulovic2012}%
  \BibitemOpen
  \bibfield  {author} {\bibinfo {author} {\bibfnamefont {G.}~\bibnamefont
  {Krstulovic}},\ }\bibfield  {title} {\bibinfo {title} {Kelvin-wave cascade
  and dissipation in low-temperature superfluid vortices},\ }\href
  {https://doi.org/10.1103/PhysRevE.86.055301} {\bibfield  {journal} {\bibinfo
  {journal} {Physical Review E}\ }\textbf {\bibinfo {volume} {86}},\ \bibinfo
  {pages} {055301(R)} (\bibinfo {year} {2012})}\BibitemShut {NoStop}%
\bibitem [{\citenamefont {Fonda}\ \emph {et~al.}(2014)\citenamefont {Fonda},
  \citenamefont {Meichle}, \citenamefont {Ouellette}, \citenamefont {Hormoz},\
  and\ \citenamefont {Lathrop}}]{Fonda2014}%
  \BibitemOpen
  \bibfield  {author} {\bibinfo {author} {\bibfnamefont {E.}~\bibnamefont
  {Fonda}}, \bibinfo {author} {\bibfnamefont {D.~P.}\ \bibnamefont {Meichle}},
  \bibinfo {author} {\bibfnamefont {N.~T.}\ \bibnamefont {Ouellette}}, \bibinfo
  {author} {\bibfnamefont {S.}~\bibnamefont {Hormoz}},\ and\ \bibinfo {author}
  {\bibfnamefont {D.~P.}\ \bibnamefont {Lathrop}},\ }\bibfield  {title}
  {\bibinfo {title} {Direct observation of {{Kelvin}} waves excited by
  quantized vortex reconnection},\ }\href
  {https://doi.org/10.1073/pnas.1312536110} {\bibfield  {journal} {\bibinfo
  {journal} {Proceedings of the National Academy of Sciences of the United
  States of America}\ }\textbf {\bibinfo {volume} {111}},\ \bibinfo {pages}
  {4707} (\bibinfo {year} {2014})}\BibitemShut {NoStop}%
\bibitem [{\citenamefont {Homann}\ \emph {et~al.}(2009)\citenamefont {Homann},
  \citenamefont {Kamps}, \citenamefont {Friedrich},\ and\ \citenamefont
  {Grauer}}]{Homann2009}%
  \BibitemOpen
  \bibfield  {author} {\bibinfo {author} {\bibfnamefont {H.}~\bibnamefont
  {Homann}}, \bibinfo {author} {\bibfnamefont {O.}~\bibnamefont {Kamps}},
  \bibinfo {author} {\bibfnamefont {R.}~\bibnamefont {Friedrich}},\ and\
  \bibinfo {author} {\bibfnamefont {R.}~\bibnamefont {Grauer}},\ }\bibfield
  {title} {\bibinfo {title} {Bridging from {{Eulerian}} to {{Lagrangian}}
  statistics in {{3D}} hydro- and magnetohydrodynamic turbulent flows},\ }\href
  {https://doi.org/10.1088/1367-2630/11/7/073020} {\bibfield  {journal}
  {\bibinfo  {journal} {New Journal of Physics}\ }\textbf {\bibinfo {volume}
  {11}},\ \bibinfo {pages} {073020} (\bibinfo {year} {2009})}\BibitemShut
  {NoStop}%
\bibitem [{\citenamefont {Bou{\'e}}\ \emph {et~al.}(2015)\citenamefont
  {Bou{\'e}}, \citenamefont {L'vov}, \citenamefont {Nagar}, \citenamefont
  {Nazarenko}, \citenamefont {Pomyalov},\ and\ \citenamefont
  {Procaccia}}]{Boue2015a}%
  \BibitemOpen
  \bibfield  {author} {\bibinfo {author} {\bibfnamefont {L.}~\bibnamefont
  {Bou{\'e}}}, \bibinfo {author} {\bibfnamefont {V.~S.}\ \bibnamefont {L'vov}},
  \bibinfo {author} {\bibfnamefont {Y.}~\bibnamefont {Nagar}}, \bibinfo
  {author} {\bibfnamefont {S.~V.}\ \bibnamefont {Nazarenko}}, \bibinfo {author}
  {\bibfnamefont {A.}~\bibnamefont {Pomyalov}},\ and\ \bibinfo {author}
  {\bibfnamefont {I.}~\bibnamefont {Procaccia}},\ }\bibfield  {title} {\bibinfo
  {title} {Energy and vorticity spectra in turbulent superfluid {{He}} 4 from
  {{T}} = 0 to {{T}} {$\lambda$}},\ }\href
  {https://doi.org/10.1103/PhysRevB.91.144501} {\bibfield  {journal} {\bibinfo
  {journal} {Physical Review B}\ }\textbf {\bibinfo {volume} {91}},\ \bibinfo
  {pages} {144501} (\bibinfo {year} {2015})}\BibitemShut {NoStop}%
\bibitem [{\citenamefont {Polanco}\ \emph
  {et~al.}(2021{\natexlab{b}})\citenamefont {Polanco}, \citenamefont
  {M{\"u}ller},\ and\ \citenamefont {Krstulovic}}]{Circulation}%
  \BibitemOpen
  \bibfield  {author} {\bibinfo {author} {\bibfnamefont {J.~I.}\ \bibnamefont
  {Polanco}}, \bibinfo {author} {\bibfnamefont {N.~P.}\ \bibnamefont
  {M{\"u}ller}},\ and\ \bibinfo {author} {\bibfnamefont {G.}~\bibnamefont
  {Krstulovic}},\ }\href@noop {} {\bibinfo {title} {Circulation.jl: {{Tools}}
  for computing velocity circulation statistics from periodic {{3D
  Navier--Stokes}} and {{Gross--Pitaevskii}} fields}} (\bibinfo {year}
  {2021}{\natexlab{b}})\BibitemShut {NoStop}%
\bibitem [{\citenamefont {Anselmet}\ \emph {et~al.}(1984)\citenamefont
  {Anselmet}, \citenamefont {Gagne}, \citenamefont {Hopfinger},\ and\
  \citenamefont {Antonia}}]{Anselmet1984}%
  \BibitemOpen
  \bibfield  {author} {\bibinfo {author} {\bibfnamefont {F.}~\bibnamefont
  {Anselmet}}, \bibinfo {author} {\bibfnamefont {Y.}~\bibnamefont {Gagne}},
  \bibinfo {author} {\bibfnamefont {E.~J.}\ \bibnamefont {Hopfinger}},\ and\
  \bibinfo {author} {\bibfnamefont {R.~A.}\ \bibnamefont {Antonia}},\
  }\bibfield  {title} {\bibinfo {title} {High-order velocity structure
  functions in turbulent shear flows},\ }\href
  {https://doi.org/10.1017/S0022112084000513} {\bibfield  {journal} {\bibinfo
  {journal} {Journal of Fluid Mechanics}\ }\textbf {\bibinfo {volume} {140}},\
  \bibinfo {pages} {63} (\bibinfo {year} {1984})}\BibitemShut {NoStop}%
\bibitem [{\citenamefont {Vinen}\ and\ \citenamefont
  {Niemela}(2002)}]{Vinen2002}%
  \BibitemOpen
  \bibfield  {author} {\bibinfo {author} {\bibfnamefont {W.~F.}\ \bibnamefont
  {Vinen}}\ and\ \bibinfo {author} {\bibfnamefont {J.~J.}\ \bibnamefont
  {Niemela}},\ }\bibfield  {title} {\bibinfo {title} {Quantum {{Turbulence}}},\
  }\href@noop {} {\bibfield  {journal} {\bibinfo  {journal} {Quantum
  Turbulence}\ ,\ \bibinfo {pages} {65}} (\bibinfo {year} {2002})}\BibitemShut
  {NoStop}%
\end{thebibliography}%

\end{document}